\documentclass[journal]{IEEEtran}
\usepackage{graphicx}
\usepackage{amsmath}
\usepackage{booktabs}
\usepackage{stfloats}
\usepackage{subfigure}
\usepackage{algpseudocode}
\usepackage{algorithm}
\usepackage{multirow}
\usepackage{algorithmicx}
\usepackage{hyperref}
\usepackage[misc]{ifsym}

\ifCLASSINFOpdf
\else
\fi

\begin{document}

\title{A Quantitative Framework for Network Resilience Evaluation using Dynamic Bayesian Network}

\author{
\IEEEauthorblockN{Shanqing~Jiang\IEEEauthorrefmark{1}\IEEEauthorrefmark{2}\IEEEauthorrefmark{3},
	Lin~Yang\IEEEauthorrefmark{2}, Guang~Cheng\IEEEauthorrefmark{1}\IEEEauthorrefmark{3}$^{(\textrm{\Letter})}$,
	Xianming~Gao\IEEEauthorrefmark{2},
	Tao~Feng\IEEEauthorrefmark{2},
	Yuyang~Zhou\IEEEauthorrefmark{1}\IEEEauthorrefmark{3}
}

\IEEEauthorblockA{\IEEEauthorrefmark{1}School of Cyber Science and Engineering, Southeast University, Nanjing, China}\\
\IEEEauthorblockA{\IEEEauthorrefmark{2}National Key Laboratory of Science and Technology on Information System Security, Beijing, China}\\
\IEEEauthorblockA{\IEEEauthorrefmark{3}Key Laboratory of Computer Network and Information Integration, Ministry of Education, Nanjing, China}

\thanks{Shanqing~Jiang is with the School of Cyber Science and Engineering, Southeast University, Nanjing, China, and also with the National Key Laboratory of Science and Technology on Information System Security, Beijing, China, e-mail: sqjiang@njnet.edu.cn}
\thanks{Lin~Yang, Xianming~Gao and Tao~Feng are with the National Key Laboratory of Science and Technology on Information System Security, Beijing, China, email: \{nudt\_lyang, nudt\_gxm, feng09\}@163.com}
\thanks{Guang~Cheng and Yuyang~Zhou are with the School of Cyber Science and Engineering, Southeast University, Nanjing, China, e-mail: \{gcheng, yyzhou\}@njnet.edu.cn}
}


\maketitle

\begin{abstract}
Measuring and evaluating network resilience has become an important aspect since the network is vulnerable to both uncertain disturbances and malicious attacks. Networked systems are often composed of many dynamic components and change over time, which makes it difficult for existing methods to access the changeable situation of network resilience. This paper establishes a novel quantitative framework for evaluating network resilience using the Dynamic Bayesian Network. The proposed framework can be used to evaluate the network’s multi-stage resilience processes when suffering various attacks and recoveries. First, we define the dynamic capacities of network components and establish the network’s five core resilience capabilities to describe the resilient networking stages including preparation, resistance, adaptation, recovery, and evolution; the five core resilience capabilities consist of rapid response capability, sustained resistance capability, continuous running capability, rapid convergence capability, and dynamic evolution capability. Then, we employ a two-time slices approach based on the Dynamic Bayesian Network to quantify five crucial performances of network resilience based on core capabilities proposed above. The proposed approach can ensure the time continuity of resilience evaluation in time-varying networks. Finally, our proposed evaluation framework is applied to different attacks and recovery conditions in typical simulations and real-world network topology. Results and comparisons with extant studies indicate that the proposed method can achieve more accurate and comprehensive evaluation and can be applied to network scenarios under various attack and recovery intensity.
\end{abstract}



\section{Introduction}
\IEEEPARstart{T}{he} computer networked system has increasingly become a critical infrastructure in supporting a wide range of services in the economy, education, and government sectors. The fact that these sectors are facing severe challenges and threats, such as human error, equipment failure, deliberate attacks, natural disasters, and economic crisis. From a time perspective, it is impossible to respond to all unknown threats in advance. From a space perspective, it is rather difficult to transfer or rebuild network infrastructure before or after a disaster occurs. Therefore, the network system should have the ability to resist internal or external threats and sustain normal services and tasks, rather than providing absolute security.

The word “resilience” comes from the Latin word “resilio”, which literally means “to bounce back”, referring to a system’s ability to return to normal condition after challenging or destructive events. This broad definition applies to fields as diverse as ecology, materials science, psychology, economics, and engineering \cite{ref1}. A variety of definitions on resilience have been proposed by researchers in multiple disciplines. For example, Pregenzer \cite{ref2} defined resilience as the “measure of a system’s ability to absorb continuous and unpredictable change and still maintain its vital functions.” Woods \cite{ref3} declared that resilience is the system’s ability to create foresight, identify risks, and mitigate risks before adverse consequences. Haimes \cite{ref4} defined resilience as “the ability of the system to withstand a major disruption within acceptable degradation parameters and to recover with an acceptable cost and suitable time.” Defined by the National Academy of Sciences (NAS) as “the ability to plan and prepare for, absorb, recover from disasters and more successfully adapt to adverse events”, resilience is becoming one of the most widely used attributes in various organizations and governments \cite{ref5}.

There also exists discussions in the literature about the confusion between resilience and other system attributes, such as robustness, vulnerability, and reliability. Robustness is usually defined as insensitivity to uncertain disturbances \cite{ref6}. Uday and Marais \cite{ref7} added that the purpose of robustness is to immediately minimize performance loss after disturbance; in contrast, resilience allows for some performance loss in the hope that performance can be restored over time. Vulnerability focuses on susceptibility to known disturbances that can be obtained by both attacker and defender in advance \cite{ref8}. Reliability refers to the system’s ability and its components to accomplish required functions within a specified time under stated conditions \cite{ref7}. Different from the above concepts of system attributes, resilience places greater emphasis on the recovery and evolutionary ability to resist unknown future threats.

There emerges diverse network attacks and threats, and network security defenses are also being developed into a proactive defense direction, such as the Moving Target Defense \cite{ref9}. Therefore, it has become an urgent problem to reasonably and effectively evaluate and improve the network resilience in various attack and defense scenarios. In general, much research on resilience has been performed from the following three aspects. Measurement and evaluation research is the first step to study network resilience, including failure models, measure indicators, and aggregation models. Second, optimized and improved strategies for network resilience are conducted under specific network scenario based on entity-related analysis. Third, research based on resilience focuses on trade-off between networking performance and resource invested in practical network, such as transportation network \cite{ref10} and power supply network \cite{ref11}. However, the existing research on network resilience lack general measurement methods and standards suitable for different network scenarios, and most of them were only used to evaluate network resilience on time-invariant network, which cannot reflect the dynamic characteristic of real-world network. In this paper, a quantitative framework for network resilience evaluation is established using the Dynamic Bayesian Network based on modeling of five core resilient capabilities. The proposed framework is suitable for time-varying network and can be used to describe the process of network resilience, including preparation, resistance, adaptation, recovery, and evolution.

In this paper, our contributions can be summarized as follows:

1)	We propose a general resilience evaluation framework to describe the multi-stage resilience transformation process of network including preparation, resistance, adaptation, recovery, and evolution.

2)	We define and describe five core capabilities of network resilience based on several basic network measurement indicators and dynamic abilities of network components.

3)	We use the Dynamic Bayesian Network to describe the time-varying resilience of network and combine it with the multi-stage network resilience transformation process to evaluate network resilience.

The rest of this paper is structured as follows. In section II, we begin with the definition of network resilience, and we propose an evaluation framework of network resilience. Evaluation framework and core capabilities indicators are proposed in section III. In section IV, the resilience time-varying process based on the Dynamic Bayesian Network are defined and modeled in sections III. In section V, we use two group of experiments to verify the rationality and performance of our proposed method. Finally, we provide the related works of network resilience evaluation in section VI and a brief conclusion and directions for future studies in section VII.

\section{Description of Network Resilience}
This section provides the definition of network resilience and the resilience evaluation framework, which describes the multi-stage resilience process of transformation and the corresponding resilient abilities. The evaluation framework serves as the methodological background for the network resilience’s core capabilities measurement and the Dynamic Bayesian Network modeling that will be discussed in section IV.
\subsection{Network Resilience Definition}
There is almost no universally accepted definition of resilience in network. However, the literature agrees on several key aspects of network resilience evaluation, which involves the probability of disruption, the impacts of those disruptions, and the behavior of recovering from the disruption to the normal state \cite{ref17}\cite{ref18}. In this paper, we enrich and complement the definition of resilience in network based on NAS’s version \cite{ref5}. The resilient network should be equipped with the resistance’s ability to inhibit or mitigate internal and external disruptions and networking attacks, the ability to adapt against disruption by adjusting network structure and functional elements, the ability to recover from low performance state to normal running state, and the ability to evolve to a more stable state by intelligently reallocating network resources. When facing future networking adverse events, networks with resilient abilities will adopt faster responses and optimal strategies that minimize network damage.

\begin{figure}[!bp]
	\centering
	\includegraphics[scale=.82]{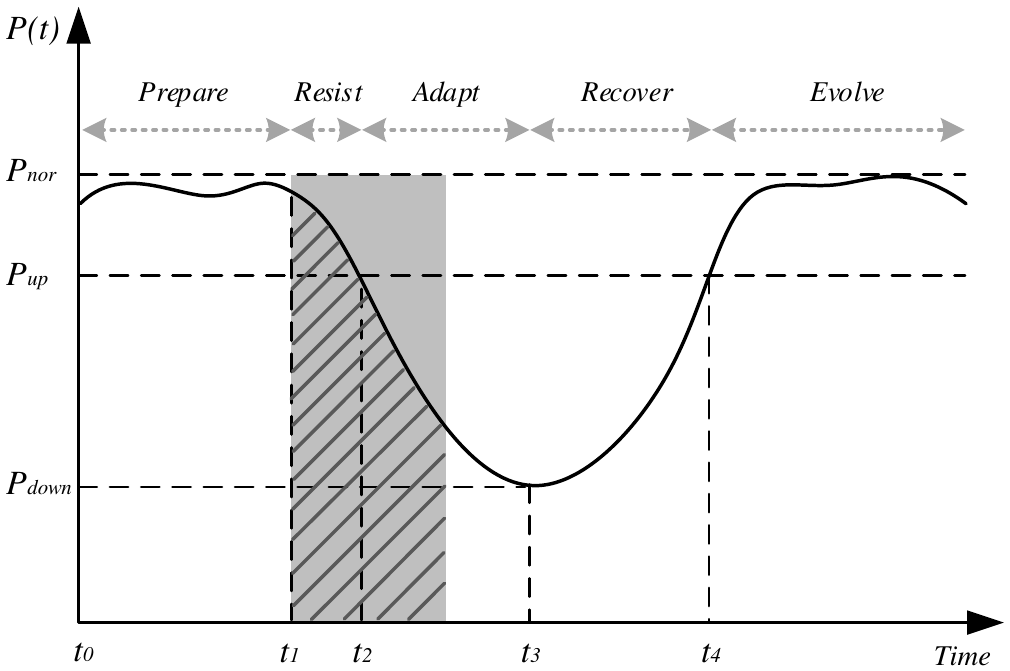}
	\caption{The time-varying process of network resilience}
	\label{fig1}
\end{figure}

\subsection{Time-varying Process of Network Resilience}
As illustrated in Fig.\ref{fig1}, a complete change process of network resilience will go through several time-varying processes of normal network fluctuation (preparation), resistance, adaptation, recovery, and evolution. A quantifiable and time-independent system performance function $ P(t) $ is the basis for evaluating the resilience of the network system \cite{ref21}. The curve in Fig.\ref{fig1} has a normal value waving around $ P_{nor} $ caused by normal network fluctuations. The network operates at this performance level until it is disrupted or attacked at time $ t_{1} $. The network system’s risk-aware component will distinguish between disruptions and normal performance fluctuations and initiate resistant approaches to prevent further degradation of network performance after $ t_{1} $. If the resistance stage is successful, the network will return to a normal operation stage. However, when the performance function $ P(t) $ declines to $ P_{up} $ at time $ t_{2} $, the network system will no longer be able to provide a stable service performance and cannot adjust itself back to a normal operation stage. Then, during the adaptation stage, the network performance drops to $ P_{down} $ at $ t_{3} $ under the adaptation strategies, including adjusting the network structure and configuration. After time $ t_{3} $, the network system launches recovery policies to improve performance until it achieves $ P_{up} $ at a later time $ t_{4} $. Since the network performance returns to the normal-level threshold, the historical information during destroy and recovery stages can be used to better optimize networking configuration strategies for improving network performance. As shown in Fig.\ref{fig1}, the area between resilience curve and dotted $ P_{nor} $, also called resilience triangle \cite{ref12}, is the total loss as a consequence of the disruption.

Let $ R(t) $ be the network resilience, which describe the instantaneous resilience performance functionality  at time $ t $, normalized by the expected network functionality $ P_{nor} $ supposing that the network has not been affected by disruption. Let $ R(T) $ be the cumulative network resilience during time $ T(T>t_{1}) $, which describe the total loss of network resilience over a period of time $ T $. The formula of $ R(t) $ and $ R(T) $ as Equation (\ref{eq1}) and (\ref{eq2}).
\begin{equation}
\label{eq1}
R(t) = \frac{{\int_t^{t + \Delta t} {P(\tau )d\tau } }}{{\int_t^{t + \Delta t} {P({t_0})d\tau } }}
\end{equation}
\begin{equation}
\label{eq2}
R(T) = \frac{{\int_{t_{1}}^{t_{1} + T} {P(\tau )d\tau } }}{{\int_{t_{1}}^{t_{1} + T} {P({t_0})d\tau } }}
\end{equation}

\begin{table*}[!bp]
	\centering
	\begin{tabular}{llll}
		\toprule
		RRC & Rapid response capability & $\textit{BN}_{i}$      & Betweenness of node \textit{i} \\
		SRC & Sustained resistance capability & $\textit{BE}_{ij}$     & Betweenness of edge \textit{(i, j)} \\
		CRC & Continuous running capability & $\textit{Bw}_{ij}(t)$  & The bandwidth of edge \textit{(i, j)} at time \textit{t} \\
		RCC & Rapid convergence capability & $\textit{Rtt}_{ij}(t)$ & The \textit{RTT} between edge \textit{(i, j)} at time \textit{t} \\
		DEC & Dynamic evolution capability & $\textit{Deg}_{i}$  & The degree of node \textit{i} \\
		MAI & The maximum of the total four abilities of each node & $ FR(G) $              & Flow robustness of network \\
		\textit{G(N, E)}       & Network with a set of nodes \textit{N}  connected by a set of edges \textit{E} & $ R(G) $ & Effective graph resistance \\
		\textit{T}             & Standard time interval & $ H(G) $ & Structure entropy of network \\
		$\textit{O}_{i}(t)$    & The observation ability of node \textit{i} at time \textit{t} & $ DP_{i} $ & The ratio of node \textit{i} degree to total node degree \\
		$\textit{C}_{i}(t)$    & The control ability of node \textit{i} at time \textit{t} & $ P_{p}(t) $ & Network performance of preparing at time \textit{t} \\
		$\textit{D}_{i}(t)$    & The decision ability of node \textit{i} at time \textit{t} & $ P_{s}(t) $ & Network performance of resisting at time \textit{t} \\
		$\textit{A}_{i}(t)$    & The action ability of node \textit{i} at time \textit{t} & $ P_{a}(t) $ & Network performance of adapting at time \textit{t} \\
		$\textit{L}_{i}$       & Likelihood of disruption on node \textit{i} & $ P_{r}(t) $ & Network performance of recovering at time \textit{t} \\
		$\textit{L}_{ij}$      & Likelihood of disruption on edge \textit{(i, j)} & $ P_{e}(t) $ & Network performance of evolving at time \textit{t} \\
		$\textit{R}_{i}$       & Repair rate function of node \textit{i} & $ T_{d} $ & The time of destroy occurring \\
		$\textit{R}_{ij}$      & Repair rate function of edge \textit{(i, j)} & $ N_{d} $ & The number of destroyed nodes \\
		$\textit{I}_{i}$       & Criticality of node \textit{i} on the network & $ N_{r} $& The number of recovery nodes at each recovery step\\
		$\textit{I}_{ij}$      & Criticality of edge \textit{(i, j)} on the network & $ PA $         & The probability of effectively attack \\
		$\hat{\tau}$           & Network criticality & $ PR $ & The recovery probability of nodes and edges \\
		\bottomrule
	\end{tabular}
\end{table*}

It can be graphically displayed that the $ R(T) $ in formula (2) focuses on the whole resilience stage and its value is in range of [0, 1], which is quantified by the shadow area's ratio below the curve to the total rectangle shadow area in Fig.\ref{fig1}. The value is influenced by the resilience process time and the minimum performance threshold $ P_{down} $. When $ t_{1}<T<t_{3} $, a slower curve descent rate and a higher minimum performance threshold will result in a larger $ R(T) $ value, meaning that the network is more resilient.

\subsection{Factors of Network Resilience}
The computer network is a system in which multiple sub-systems with independent functions are connected by communication links; these links are managed by a network operating system and protocol software to realize data communication and network resource sharing. The multiple subsystems with independent functions can be abstracted as nentwork nodes with different capabilities. For example, the Intrusion Detection System can be regarded as a node with observation capacity, and the controller cluster can be considered as a node with control capacity. Moreover, a data center has powerful decision capacity, and massive network terminal devices are equipped with special action capacity. By analyzing the functional characteristics of actual network units, inspired by the “OODA” loop \cite{ref20}, we define four kinds of variable capacities of network nodes, including observation capacity, control capacity, decision capacity, and action capacity. Additionally, adjustable bandwidth and delay are enabled to network links, which also refer to the actual network system. These dynamic properties mentioned above provide basic motivation for network resilience evaluation and are also the internal perspective for network resilience. Referring to numerous quantitative studies in the literature \cite{ref19}, this paper considers both the internal perspective and external perspective of network system. It is precisely because of the internal networking elements’ specific functions that the performance of network resilience from external perspective can be established, including preparation, resistance, adaptation, recovery, and evolution.

The above four node's capacities and two edge's capacities will have significant effects on the network resilience derived from the proposed framework. Once the network system is attacked, dynamically and intelligently tuning these basic capabilities can help achieve the network's resilience. Our starting point of the proposed framework is a computer network system under the environment of cyberattacks. If researcheres want to apply the evaluation model to real-world network such as smart power grids and transportation networks, they need to make appropriate capacity mapping for the nodes and edges firstly. For example, in a smart power grid, the intrusion detection equipment is a node with observation capacity; the safety isolation device is a node with action capacity; the computing platform is a node with decision capacity; the monitoring center is a node with control capacity. And these capacities can also be adjuested over time according to the real-time requirements of network system. The specific application to actual networks requires further clarification and functional refinement of the elements in the network.

\section{Framework and Methodology}
\begin{figure*}[!bp]
	\centering
	\includegraphics[scale=1.1]{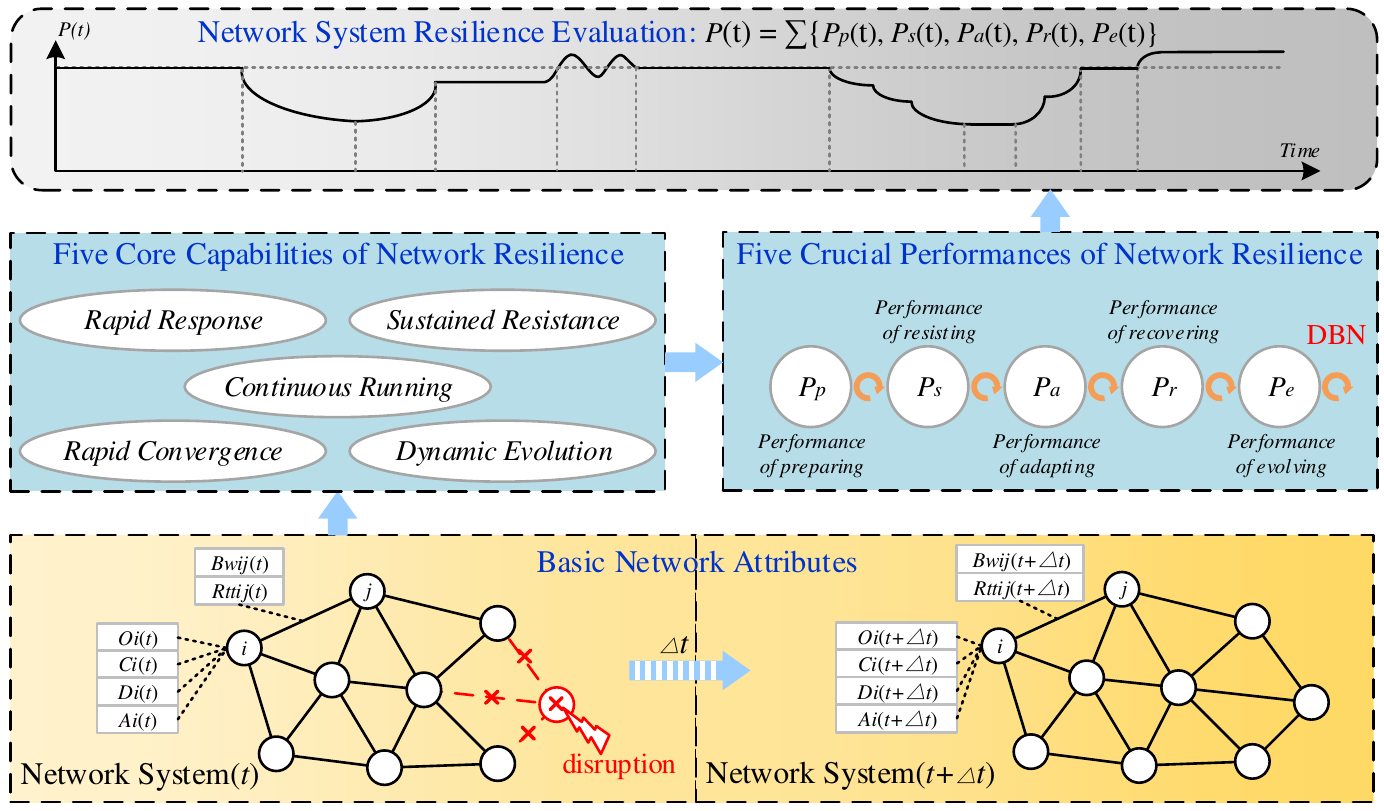}
	\caption{Network resilience Evaluation Framework: based on the measurement of dynamic basic network attributes and probabilities in time-varying network, five defined core capabilities are calculated to reflect the temporal multidimensional resilient capability, then five crucial performances of network resilience are evaluated based on DBN (Dynamic Bayesian Network) to reveal the time continuous resilience of network.}
	\label{fig2}
\end{figure*}

This section begins with the introduction of the network’s basic attributes, including graph spectral matrices and time-varying capacities of network elements. Based on these attributes, we build the measurement of five core capabilities during the resilience process as mentioned in section II. Next a general modeling framework of resilience evaluation based on the Dynamic Bayesian Network is established to quantify resilience in a time-varying network. The evaluation framework of network resilience is illustrated as Fig.\ref{fig2}. The notations are displayed in the table below to define the network attributes and probability parameters of network components during evaluation.

\subsection{Basic Networking Attributes}
The mathematical model for the network resilience evaluation concerns a basic network $ G(N, E) $, which comprises a set of nodes $ N $ connected by a set of edges or arcs $ E $. The relationship between nodes and edges in a network can be visually described by graph. Several graph spectral matrices, such as \textit{algebraic connectivity}, \textit{natural connectivity}, and \textit{flow robustness}, are generally employed to measure the robustness and resilience of network in generally \cite{ref22}. The graph's topology $ G(N, E) $ can be represented by adjacency matrix and Laplacian matrix. Let $ \{\lambda_{1}, \lambda_{2}, ..., \lambda_{n}\} $ represent the eigenvalues list of the Laplacian matrix. And all the variables defined for network will be normalized to value in the interval [0,1] when calculating following equations.

\textit{Flow robustness}, denoted as \textit{FR(G)}, is a graph metric that measures the ratio of the number of available flows to the number of total flows in the network \cite{ref22}. A flow is considered available if at least one of its paths remains reachable after link or node failures. The number of total flows represents the maximum of network flows. For example, a connected network with $ \left| N \right| $ nodes has $ \left| N \right|(\left| N \right|-1)/2 $ flows between all node pairs. The range of flow robustness values is between 0 and 1, where 1 means that the graph is a completely connected graph, and 0 indicates that the nodes cannot communicate with each other in the whole network. Let $ \{C_{i};1<i<k\} $ be the set of connected sub-graph in given network $ G(N, E) $, and the $ C_{i} $ of a network can be calculated by union-find set within linear time complexity. The union-find set is a tree-shaped merging and searching data structure, which can solve the problem of a disjointed search with constant-level time and space consumption. It will not consume too much computing resources when calculating the connected subgraph even in large-scale network. The flow robustness can be calculated as Equation (\ref{eq3}).
\begin{equation}
\label{eq3}
FR(G) = \frac{{\sum\nolimits_{i = 1}^k {\left| {{C_i}} \right|(\left| {{C_i}} \right| - 1)} }}{{\left| N \right|(\left| N \right| - 1)}},\quad 1 \le k \le \left| N \right|
\end{equation}

\textit{Effective graph resistance}, denoted as \textit{R(G)}, is a graph metric that measures the network's resistance against nodes or edges destroyed \cite{ref23}. The normalized \textit{R(G)} is calculated as Equation (\ref{eq4}), where $ \lambda _i $ is the non-zero eigenvalue of the given graph’s Laplacian matrix $ L $, where the values of $ R{(G)^*} $ lie in the interval $ [0, 1] $.
\begin{equation}
\label{eq4}
R{(G)^*} = \frac{{\left| N \right| - 1}}{{\left| N \right|\sum\nolimits_{i = 2}^{N} {\frac{1}{{{\lambda _i}}}} }}
\end{equation}

Besides the measurement of graph spectral matrices, the time-varying attributes of network elements mentioned in section II-A are also crucial factors on modeling core capability of network resilience. With the development of the Software Defined Network (SDN) and Network Function Virtualization (NFV) technologies, the network become more dynamically reconfigurable and programmable \cite{ref25}. Moreover, network delay and bandwidth can be uniformly scheduled by routers with programmable kernel to achieve the network's modifiability and controllability\cite{ref26}. As described in section II-C, different network sub-systems and elements equip different networking capacities, which can be more convenient with a wide deployment of SDN and NFV, and the capacities of network elements can be dynamically adjusted over time according to the needs of application scenarios. In order to reflect the dynamic characteristics of network elements in the process of measuring network resilience, we simplify and refine the key time-varying capacities of nodes and edges in this paper. The nodes in the network will be equipped with four kinds of capacities, including observation capacity, control capacity, decision capacity and action capacity, which can be adjusted over time, and the summation is within a certain range, considering that the resource and computing power of nodes are limited in real network. We define the maximum of the total four capacities of each node as \textit{MAI}. The capacity of node is denoted as $ Cap(N) $, described as Equation (\ref{eq5}).
\begin{equation}
\label{eq5}
\begin{split}
Cap(N) = \{  < {O_i}(t),{C_i}(t),{D_i}(t),{A_i}(t) > ;i \in N\}\\
0 \le {O_i}(t)+{C_i}(t)+{D_i}(t)+{A_i}(t) \le \textit{MAI};i \in N\
\end{split}
\end{equation}
Meanwhile, the edges' capacities in network will be measured by the maximum bandwidth, real-time bandwidth and RTT latency, which are established on the adjacency matrix of network. Define $ \textbf{\textit{MaxBw}}={\left( {M{Bw(t)_{ij}}} \right)_{N \times N}} $ as the maximum bandwidth matrix, and $ \textbf{\textit{Bw}}={\left( {{Bw(t)_{ij}}} \right)_{N \times N}} $ as the real-time bandwidth matrix, and $ \textbf{\textit{Rtt}}={\left( {{Rtt(t)_{ij}}} \right)_{N \times N}} $ as the RTT latency matrix.
 
\subsection{Core Capabilities of Resilience}
In our previous definition, the resilient network should be provided with the resisting ability to mitigate or prevent internal and external disruptions and networking attacks, the ability to adapt against disruption by adjusting network structure and functional elements, the ability to recover from low performance state to normal running state, and the ability to evolve to a more stable state by intelligently reallocating network resources.

Meanwhile, a complete network resilient process should go through the following stages: preparing, resisting, adapting, recovering, and evolving. There are subtle differences between the above two perspectives. The former refers to resilience as a type of overall network capability, similar to CIA principles in information security system \cite{ref27}. The latter perspective, however, considers resilience as a manifestation or performance of network during operation process. Fundamentally, this is the network's resilient capability that results in resilience performance. Therefore, five core capabilities of resilience will be defined and modeled to describe the resilience capability more specifically as follows. It should be noted that network resilience is established on time-varying dimensions; therefore, the following five capabilities are transient capabilities, which can vary in values over time.

\subsubsection{Rapid Response Capability}
The rapid response capability (RRC) is defined as the system's response speed and emergency capability against disturbance or cyber-attacks. The system can take emergency rescue and recovery measures in earlier times when it has better rapid response capability. This capability is related to the perceptual or observing ability of network components (nodes) and the transmission ability of network links. We combine the abilities of nodes and links with graph theory. In a definite scale network, the node with larger observing ability and more connected edges will has more observation capability, and the link with more bandwidth and less transmission delay will has more rapid response capability. Therefore, the network nodes' observing ability is defined as the product of observation ability and degree distribution of network nodes. The links' transmission ability can be determined by the ratio of edges' betweenness centrality and the RTT delay between node pairs. The formula is established as Equation (\ref{eq6}), and the calculation process only considers the value of each parameter, so there is no exact unit for RRC.
\begin{equation}
\label{eq6}
RRC = \frac{1}{{\left| N \right|}}\sum\limits_{i = 1}^N {De{g_i}{O_i}}  + \frac{1}{{\left| E \right|}}\sum\limits_{ij = 1}^E \frac{B{E_{ij}}}{{Rt{t_{ij}}}}
\end{equation}

\subsubsection{Sustained Resistance Capability}
The sustained resistance capability (SRC) is defined as the system’s ability to prevent a rapid decline in network performance, which is related to the resources’ redundancy, the network topology’s robustness, the regional network’s autonomous intelligent management, and how to prevent the cascading failures’ propagation. The SRC capability can directly affect the resistance duration and the network performance’s minimum limit. The formula of SRC can be described as Equation (\ref{eq7}). The numerator considers the average effective gragh resistance from a graph theory point of view, and the denominator represents the harm caused by the destruction of network elements, which is contributed by the criticality and disruption likelihood of nodes and edges together.
\begin{equation}
\label{eq7}
SRC = \frac{{R{{(G)}^*} \times \frac{1}{{\left| N \right|}}\sum\limits_{i = 1}^N {De{g_i}} }}{{\sum\limits_{i = 1}^N {{I_i}{L_i} + \sum\limits_{ij = 1}^E {{I_{ij}}{L_{ij}}} } }}
\end{equation}
where $ R{(G)^*} $ is the effective graph resistance of network $ G(N, E) $, and $ Deg_{i} $ is the degree of node $ i $, $ L_{i} $ and $ L_{ij} $ are the disruption likelihood of node $ i $ and edge $ (i, j) $, respectively. $ I_{i} $ and $ I_{ij} $ are the criticality of node $ i $ and edge $ (i, j) $ to network, respectively. Among them, the $ I_{i} $ is calculated as the product of node's betweenness and the sum of node's observation and action ability, and the $ I_{ij} $ is calculated as the product of edge's betweenness and the sum of two side node's network criticality, which can be illustrated as Equation (\ref{eq8}) and (\ref{eq9}).
\begin{equation}
\label{eq8}
{I_i} = B{N_i} \times \left( {{O_i} + {A_i}} \right),i \in N
\end{equation}
\begin{equation}
	\label{eq9}
{I_{ij}} = B{E_{ij}} \times \left( {{I_i} + {I_j}} \right),(i,j) \in E
\end{equation}

\subsubsection{Continuous Running Capability}
The continuous running capability (CRC) is defined as the system's ability for ensuring the continuous operation of the network service during low efficiency phase. The network can reduce service performance to provide less quality of service while ensuring current network security. If some forwarding nodes on the shortest path fail, the transmission task can still be completed by rerouting, despite increasing the link transmission delay. The CRC is determined by the network's flow robustness, real-time bandwidth, and edges' criticality. It can be clearly analyzed that the network can make more rerouting decision with larger $ FR(G) $, and the critical edge should be equipped with larger real-time bandwidth. The formula of CRC is illustrated as Equation (\ref{eq10}).
\begin{equation}
\label{eq10}
CRC = \frac{1}{{\left| N \right|}}FR(G) \times \sum\limits_{ij = 1}^E {{B{w_{ij}}}{I_{ij}}}
\end{equation}

\subsubsection{Rapid Convergence Capability}
The rapid convergence capability (RCC) is defined as the capability to ensure the network's rapid convergence and restoration in the recovery stage, including network status monitoring, recovery strategy deployment, etc., which involve the adjustment of node control ability, decision ability and action ability. The purpose of constructing rapid convergence ability is to speed the recovery rate and reduce the recovery-stage time. The RCC is determined by repair rate of node and edge, the node's control, decision and action ability, the edge's real-time bandwidth and RTT delay. And the formula of RCC is illustrated as Equation (\ref{eq11}). In the recovery stage, the nodes with strong control and decision abilities will have more positive impact on the network through control path, which result in the $ C_i $ and $ D_i $ are weighted by betweenness $ B{N_i} $. The node's action ability will directly affect its connected nodes, so the $ A_i $ is weighted by node degree $ De{g_i} $. By contrast, the edge's influence on RCC can be better understood that the edge with larger betweenness $ B{E_{ij}} $, larger real-time bandwidth $ B{w_{ij}} $ and less transmission delay $ Rt{t_{ij}} $ will make more positive effect on RCC. It should be noted that all variables are normalized to value in the interval [0,1], therefore the calculation of equation can make sense.
\begin{equation}
\label{eq11}
\begin{split}
RCC = \frac{1}{{\left| N \right|}}\sum\limits_{i = 1}^N {{R_i}\left[ {B{N_i}\left( {{C_i} + {D_i}} \right) + De{g_i}{A_i}} \right]}\\
+\frac{2}{{\left| N \right|\left| {N - 1} \right|}}\sum\limits_{ij = 1}^E {{R_{ij}}} B{E_{ij}} \frac{B{w_{ij}}}{Rt{t_{ij}}}
\end{split}
\end{equation}

\subsubsection{Dynamic Evolution Capability}
The dynamic evolution capability (DEC) is defined as the system's capability to continue to evolve and regenerate after recovering from damage. It involves the deep learning and ratiocination of historical destruction and recovery measures. This ability will directly depend on the network's structural entropy and the network components' maximum capability. Network with high dynamic evolution capability will be more resistant against similar destruction in the future. The network with larger structural entropy will have greater structural stability for network's dynamic evolution. Moreover, the adjustment of various abilities of nodes is within the range of the maximum resource of each node $ MA{I_i} $. And the maximum of edge's bandwidth $ MB{w_{ij}} $ is also the key factor of dynamic evolution capability. In this paper, the DEC is calculated as the product of the network's structural entropy, the maximum of the node's abilities and the maximum of the edge's bandwidth. The formula is illustrated as Equation (\ref{eq12}).
\begin{equation}
\label{eq12}
DEC = \frac{1}{{\left| N \right|\left| E \right|}}H(G)\sum\limits_{i = 1}^N {MA{I_i}} \sum\limits_{ij = 1}^E {MB{w_{ij}}}
\end{equation}
where $ H(G) $ is the network's structural entropy, which can be calculated as Equation (\ref{eq13}), where $ D{P_i} $ represents the ratio of node \textit{i} degree to total node degree. Besides, \textit{MAI} and \textit{MBw} represent the maximum of the nodes' total four capacities and the maximum of the edges' bandwidth, respectively.
\begin{equation}
\label{eq13}
\begin{split}
H(G) &=  - \sum\limits_{i = 1}^N {D{P_i}} \ln D{P_i}\\
D{P_i} &= \frac{{De{g_i}}}{{\sum\limits_{j = 1}^N {De{g_j}} }},i \in [1,N]
\end{split}
\end{equation}

\section{Resilience Evaluation Model Based on DBN}
On the basis of detailed description of the core resilience capabilities in the above section, an evaluation model suitable for the time-varying process on network resilience needs to be established. There are many modeling methods that can visually describe the cause-effect relationship in networks, which have been used to evaluate resilience, as outlined in related works. In particular, the Bayesian Network and the Dynamic Bayesian Network can capture the conditional independence between random variables, which will be more suitable for evaluating network resilience\cite{ref28}.

The Bayesian Network (BN), also known as the Belief Networks or Causal Networks \cite{ref29}, is a directed acyclic graphical pattern to describe the conditional probability relationship between data variables based on probabilistic inference theory. The nodes in BN represent random variables, and the links between them represent conditional dependencies among the variables with parent nodes, which are ruled by the conditional probability tables (CPT). However, static BN cannot be used to model time-varying system. Therefore, the Dynamic Bayesian Network (DBN) was proposed based on the hidden Markov model to satisfy the temporal system. The DBN is also called a two-time slice BN, because there are two time slices provided in DBN modeling, time slice $ t $ and $ t + \Delta t $. The discrete time slice $ \Delta t $ is usually set to 1. By dividing a time duration into a series of time slice, the DBN allows the node attribute variable $ X^{t + \Delta t}$ at time slice $ t + \Delta t $ to be conditionally dependent upon its parent nodes $ Xp_i^{t + \Delta t} $ at the same time slice, as well as its parents $ Xp_i^t $ and its own states $ X_i^t $ at the previous time slice $ t $ \cite{ref16}. The probability function of node $ X $ at the time slice $ t + \Delta t $ can be mathematically described as Equation (\ref{eq14}), where $ M $ is the number of parent nodes of node $ X $.
\begin{equation}
\label{eq14}
\begin{split}
P({X^{t + \Delta t}}) = &\prod\limits_{i = 1}^M {\left[ {P({X^{t + \Delta t}}|Xp_i^t)P({X^{t + \Delta t}}|Xp_i^{t + \Delta t})} \right]}\\
&*P({X^{t + \Delta t}}|{X^t})
\end{split}
\end{equation}
\begin{figure}[b]
	\centering
	\includegraphics[scale=.6]{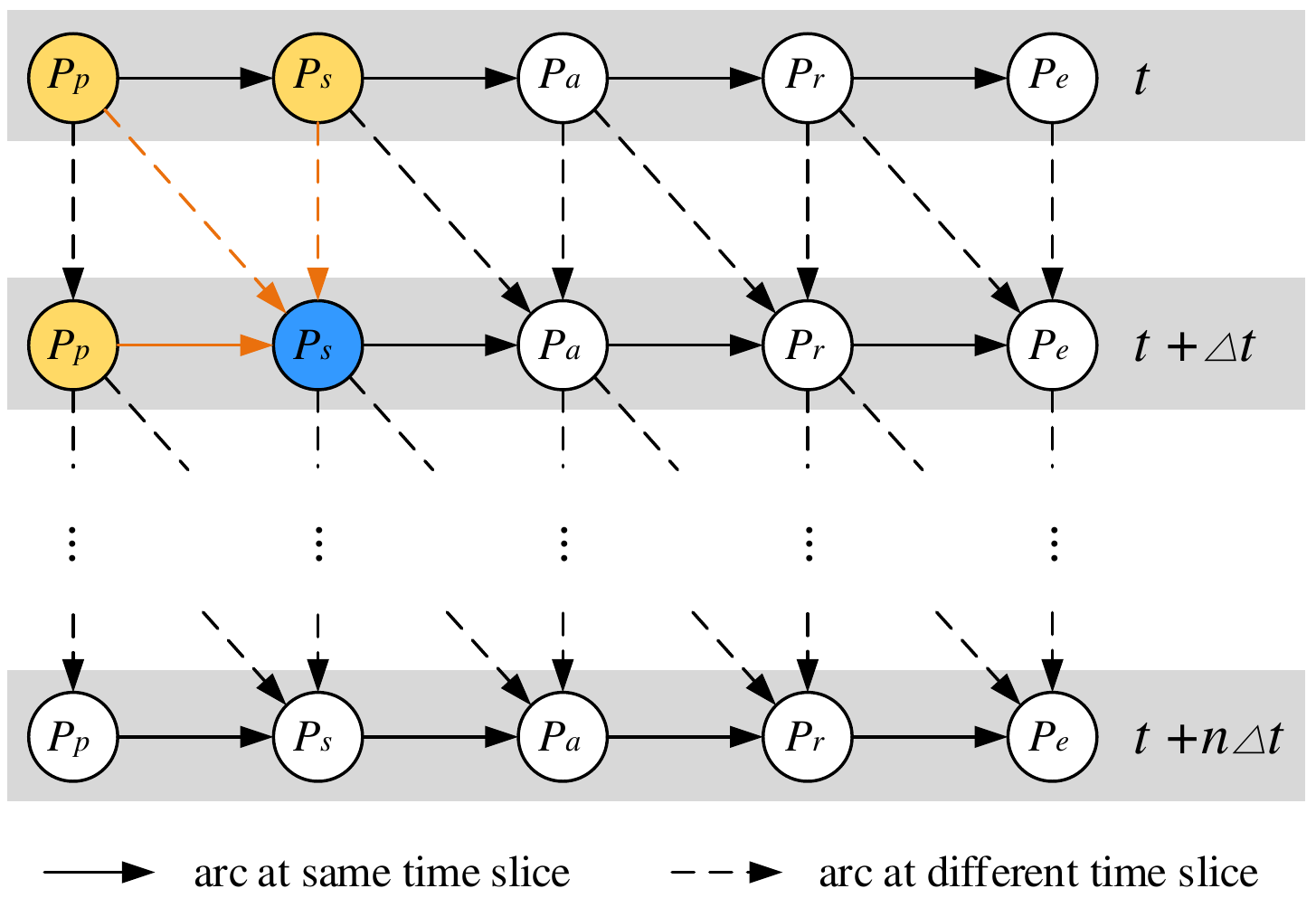}
	\caption{The DBN structure of network resilience performances}
	\label{fig3}
\end{figure}

To develop a more detailed resilience evaluation model, resilience-network characteristics rather than actual network nodes will be modeled as the nodes of DBN model. It can be seen that $ P(t) $ is a crucial characteristic when evaluating network resilience as introduced in section II-B. There is insufficient literature that formally describes $ P(t) $, or designates a simple system attribute as the quantitative standard of $ P(t) $. In this paper, we establish a detailed quantitative indicator of $ P(t) $, which is determined by five defined time-independent network performances: performance of preparing $ P_{p}(t) $, performance of resisting $ P_{s}(t) $, performance of adapting $ P_{a}(t) $, performance of recovering $ P_{r}(t) $, and performance of evolving $ P_{e}(t) $. These five performances are directly influenced directly by the five core resilience capabilities described in section III-B. Meanwhile, there also exists time dimensional interaction between these five performances based on conditional probability. For example, network A and network B equip the same sustained resistance capability (SRC) at time $ t + \Delta t $ when suffering the same destruction at time $ t $, but the performance of preparing $ P_{p}(t) $ of network A is lower than network B at time $ t $. It can be expected that network A has a better resisting performance $ P_{s}(t + \Delta t) $ than network B due to the effect of previous moment.

The DBN is adopted in modeling and evaluating network resilience in the following details, and the basic structure of DBN is shown in Fig.\ref{fig3}. First, each row of nodes represents the five resilience performances at the same time slice, where the five attribute nodes are linked by solid arcs. The solid arcs represent the conditional transition probability between the parent node and self node. Second, each column of nodes represents the time-varying states of every resilience performance. The dotted arcs between column nodes represent the temporal conditional transition probability of every resilience performance. Third, the diagonal dotted arcs represent the conditional transition probability between the parent node at the previous time slice and the self node at the current time slice. For example, the performance of resisting $ P_{s}(t) $ shown as blue node in Fig.\ref{fig3} can be calculated as Equation (\ref{eq15}).
\begin{equation}
\label{eq15}
{P_s}^{t + \Delta t} = P({P_s}^t)P({P_p}^t)P({P_p}^{t + \Delta t}) \times SR{C^{t + \Delta t}}
\end{equation}
Where the three conditional transition probability $ P({P_s}^t) $, $ P({P_p}^t) $, and $ P({P_p}^{t + \Delta t}) $ are determined by the state of three front nodes, respectively. Based on the evaluation of five resilience performances, the network resilience performance $ P(t) $ can be calculated as a weighted sum of these five performances, which is illustrated as Equation (\ref{eq16}). Finally, we can determine the network resilience by putting $ P(t) $ into the Equation(\ref{eq1}) and (\ref{eq2}). Based on the modeling above, the procedure of evaluating network resilience is summarized in Algorithm\ref{alg::algorithm1}.
\begin{equation}
\label{eq16}
\begin{split}
P(t) = &{\zeta _1}P_p^t + {\zeta _2}P_s^t + {\zeta _3}P_a^t + {\zeta _4}P_r^t + {\zeta _5}P_e^t\\
&{\zeta _1} + {\zeta _2} + {\zeta _3} + {\zeta _4} + {\zeta _5} = 1
\end{split}
\end{equation}

\begin{algorithm}[!tp]
	\caption{Network Resilience Evaluation Algorithm}
	\label{alg::algorithm1}
	\begin{algorithmic}[1]
		\Require
		Network $ G(N, E) $,
		Capacities of nodes and edges,
		$ T_{d} $,
		$ N_{d} $,
		$ PA $,
		$ N_{r} $,
		$ PR $
		\Ensure
		The resilience performance of network and the recording of time-varying resilience process
		\State Initial network with the input parameters
		\State Calculate $ RRC $, $ SRC $, $ CRC $, $ RCC $, $ DEC $ by equation (7)(8)(11)(12)(13)
		\State Calculate $ P_{p} $, $ R_{s} $, $ P_{a} $, $ P_{r} $, $ P_{e} $ by equation (16)
		\State Evaluate current resilience performance by equation (17)
		\State Set $ stepflag $ = 0
		\While {$ stepflag \leq T_{d} + \lceil N_{d}/N_{r} \rceil$}
		\If {$ stepflag = T_{d} $}
		\State Destroy the network in specified attack mode
		\EndIf
		\If {$ stepflag > T_{d} $}
		\State Resist and adjust network by modify the capacities of nodes and edges
		\State Recover network based on repair probability of components, $ R_{i},R_{ij} $, and recover speed $ N_{r} $
		\EndIf
		\State Redo step 2 to 4 to evaluate network resilience
		\State $ stepflag $ $\leftarrow$ $ stepflag + 1 $
		\EndWhile
		\State Determining the resilience of the network and calculate the accumulative resilience performance
	\end{algorithmic}
\end{algorithm}

\section{Numerical Experiment and Discussion}
In this section, we conducted two experiments on three network topologies, ER network, BA network, and one of real-world topologies Geant2012, from The Internet Topology Zoo \cite{ref30}. In the first experiment, comparative experiments were performed among our proposed evaluation approach and other network resilience measurements in existing research. In the second experiment, variable parameters were set to observe the performance of the proposed evaluation model. The simulation experiments are conducted by Python 3.8, and each network model is encapsulated as a class. During a simulation, a network object is generated and various parameters are assigned. With the execution of simulated attack and recovery, network parameters and the resilience performance of network are calculated and recorded in CSV files.

\subsection{Dataset}
In order to prove the rationality and applicability of the proposed evaluation approach, ER network, BA network, and a real-world network are selected as the network topology datasets. In ER network $ G_{ER}(n, p) $, all the nodes are randomly connected according to probability $ p $, and the degree of any node is subject to binomial distribution. The BA network $ G_{BA}(n, m) $ is generated with the rule that each newly added node is connected to $ m $ existing nodes with largest degree in the network. This happens in many real-world network models. In our experiments, the Geant2012 topology (with 40 nodes, 61 edges) was selected as an example of real-world network from the Internet Topology Zoo. ER network is set as $ G_{ER}(100, 0.3) $ and BA network is set as $ G_{BA}(100, 5) $.The size of generated network is set as 100 to keep the similar dimension with Geant2012.

\begin{figure*}[bp]
	\centering
	
	\subfigure[BA network under centrality attack]{
		\begin{minipage}[t]{0.31\linewidth}
			\centering
			\includegraphics[scale=.4]{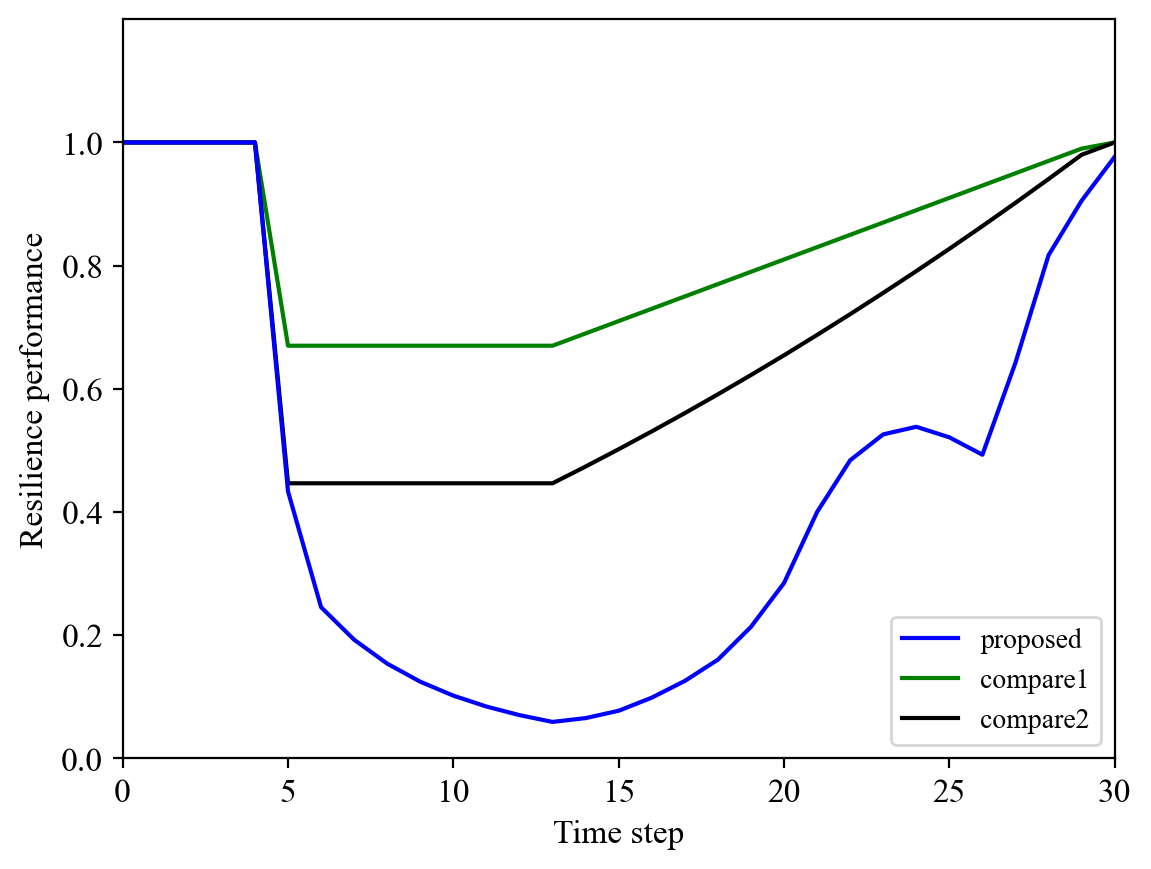}\label{fig:4a}
		\end{minipage}
	}
	\subfigure[ER network under centrality attack]{
		\begin{minipage}[t]{0.31\linewidth}
			\centering
			\includegraphics[scale=.4]{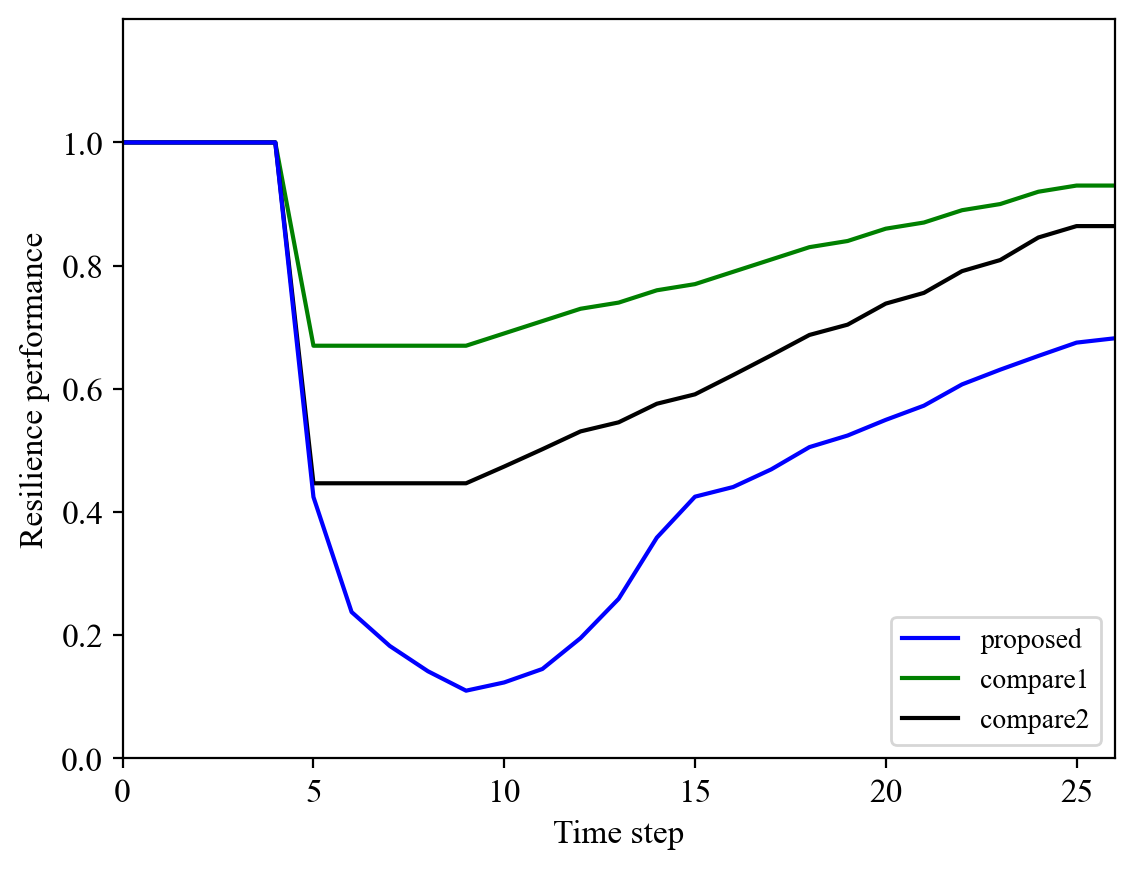}\label{fig:4b}
		\end{minipage}
	}
	\subfigure[Geant2012 network under centrality attack]{
		\begin{minipage}[t]{0.31\linewidth}
			\centering
			\includegraphics[scale=.4]{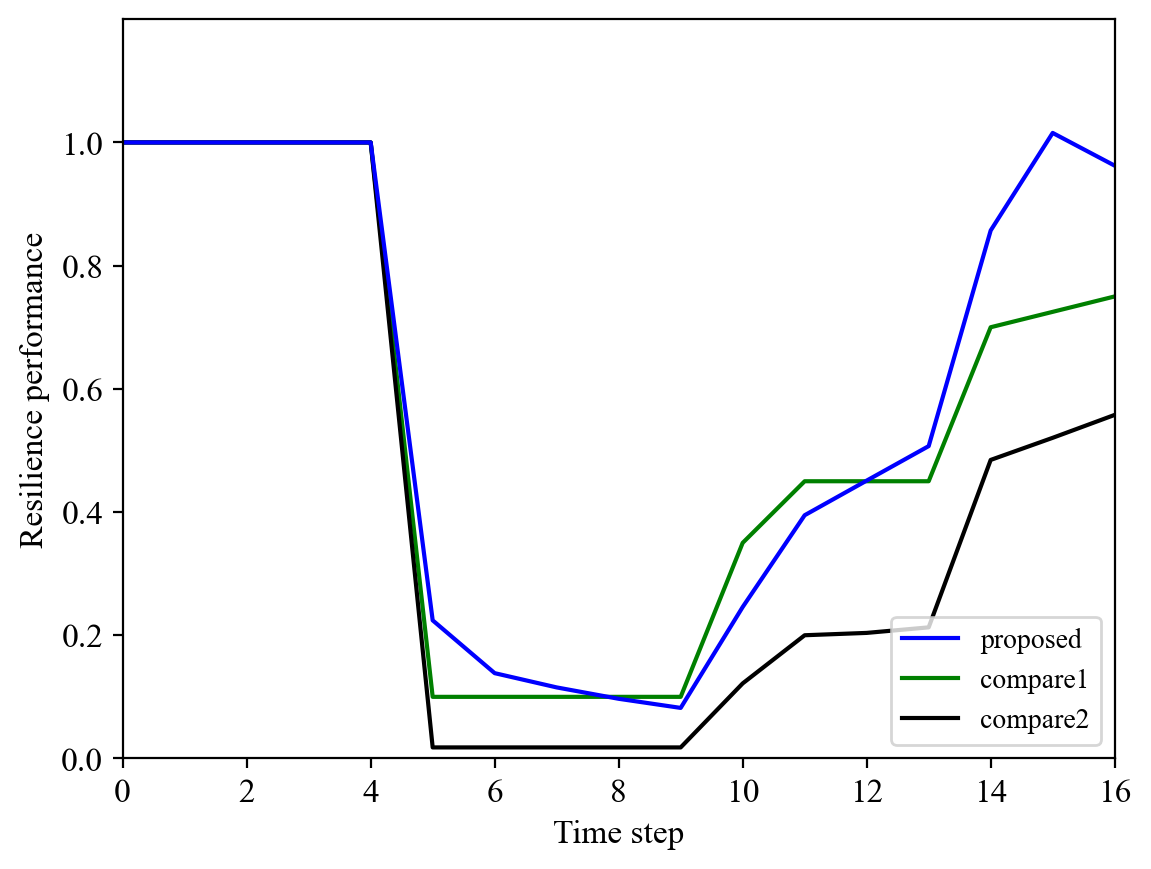}\label{fig:4c}
		\end{minipage}
	}
	
	\subfigure[BA network under random attack]{
		\begin{minipage}[t]{0.31\linewidth}
			\centering
			\includegraphics[scale=.4]{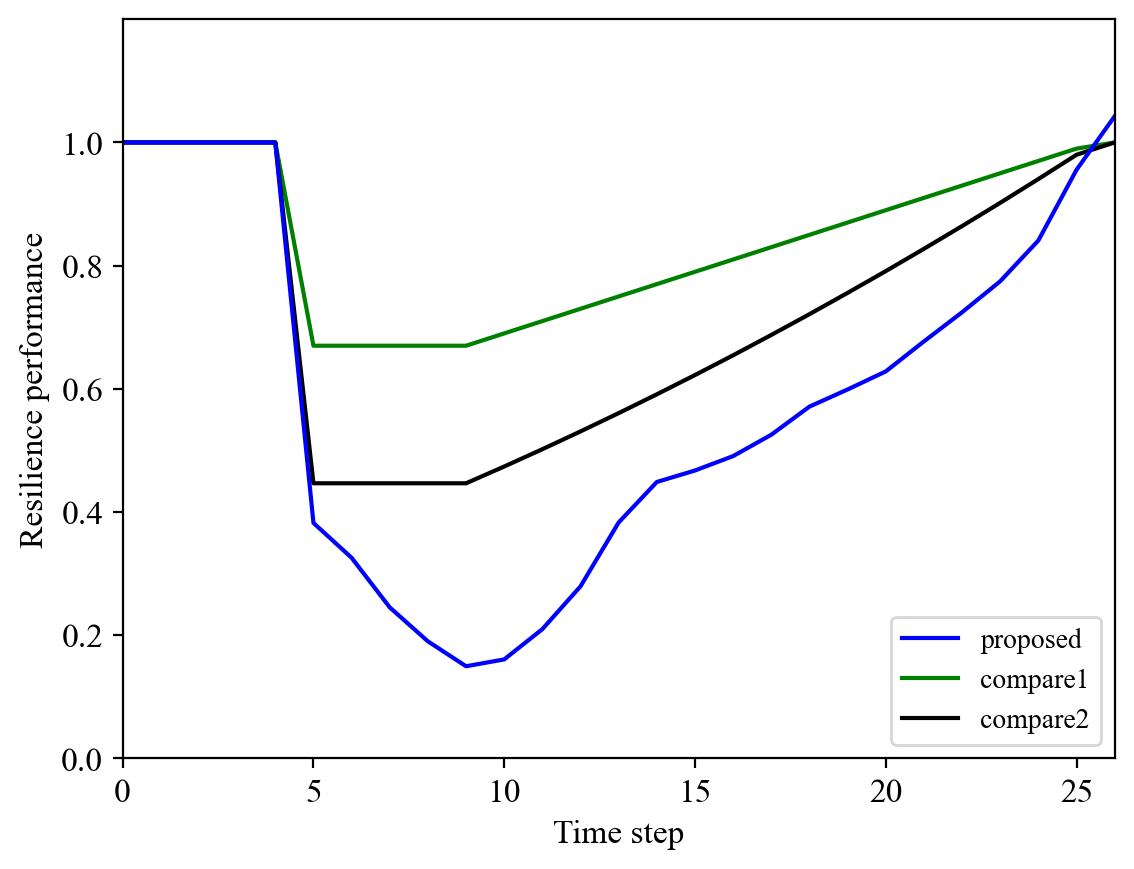}\label{fig:4d}
		\end{minipage}
	}
	\subfigure[ER network under random attack]{
		\begin{minipage}[t]{0.31\linewidth}
			\centering
			\includegraphics[scale=.4]{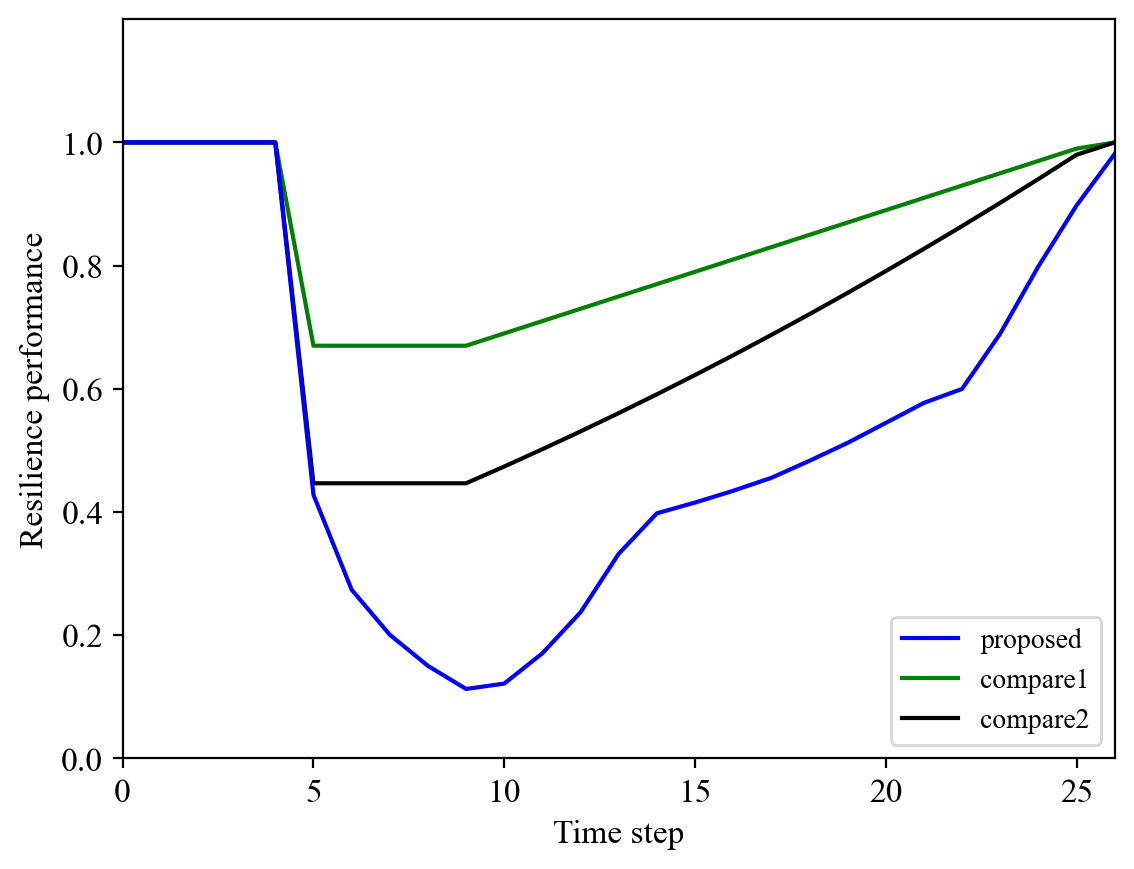}\label{fig:4e}
		\end{minipage}
	}
	\subfigure[Geant2012 network under random attack]{
		\begin{minipage}[t]{0.31\linewidth}
			\centering
			\includegraphics[scale=.4]{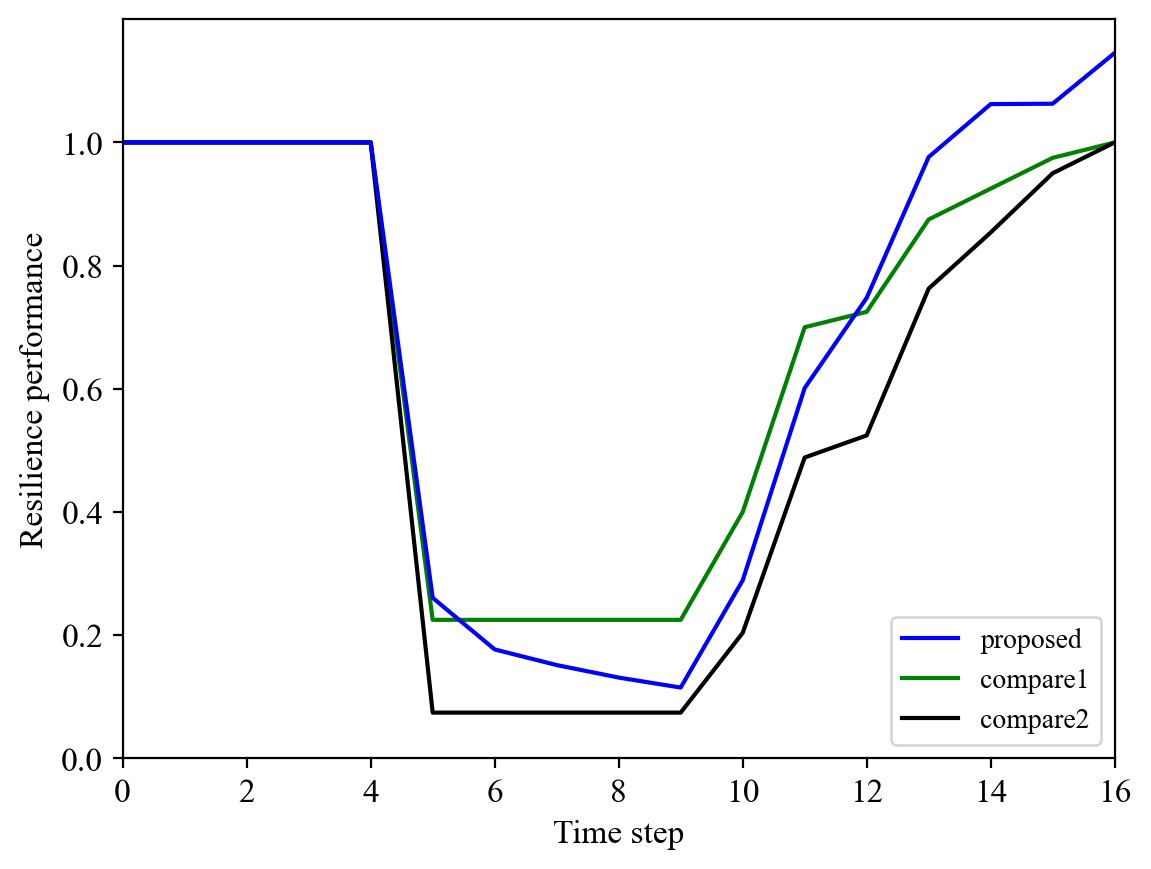}\label{fig:4f}
		\end{minipage}
	}
	
	\centering
	\caption{Comparison among the proposed approach and other approaches in three types network under centrality and random attacks}
	\label{fig4}
\end{figure*}

\subsection{Comparison among proposed and other approaches}
\begin{figure*}[hbp]
	\centering
	\subfigure[variation process of five core capabilities]{
		\begin{minipage}[t]{0.45\linewidth}
			\centering
			\includegraphics[scale=.54]{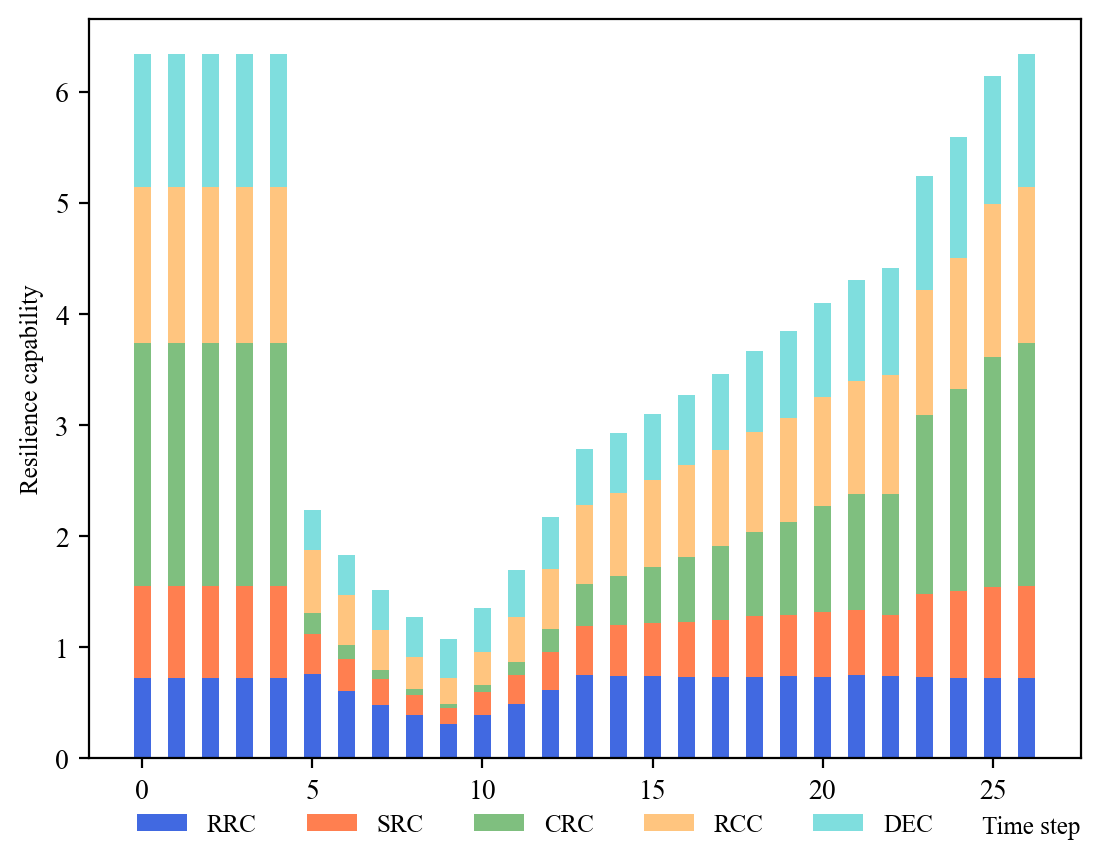}\label{fig:5a}
		\end{minipage}
	}
	\subfigure[variation process of five basic resilience performances]{
		\begin{minipage}[t]{0.45\linewidth}
			\centering
			\includegraphics[scale=.54]{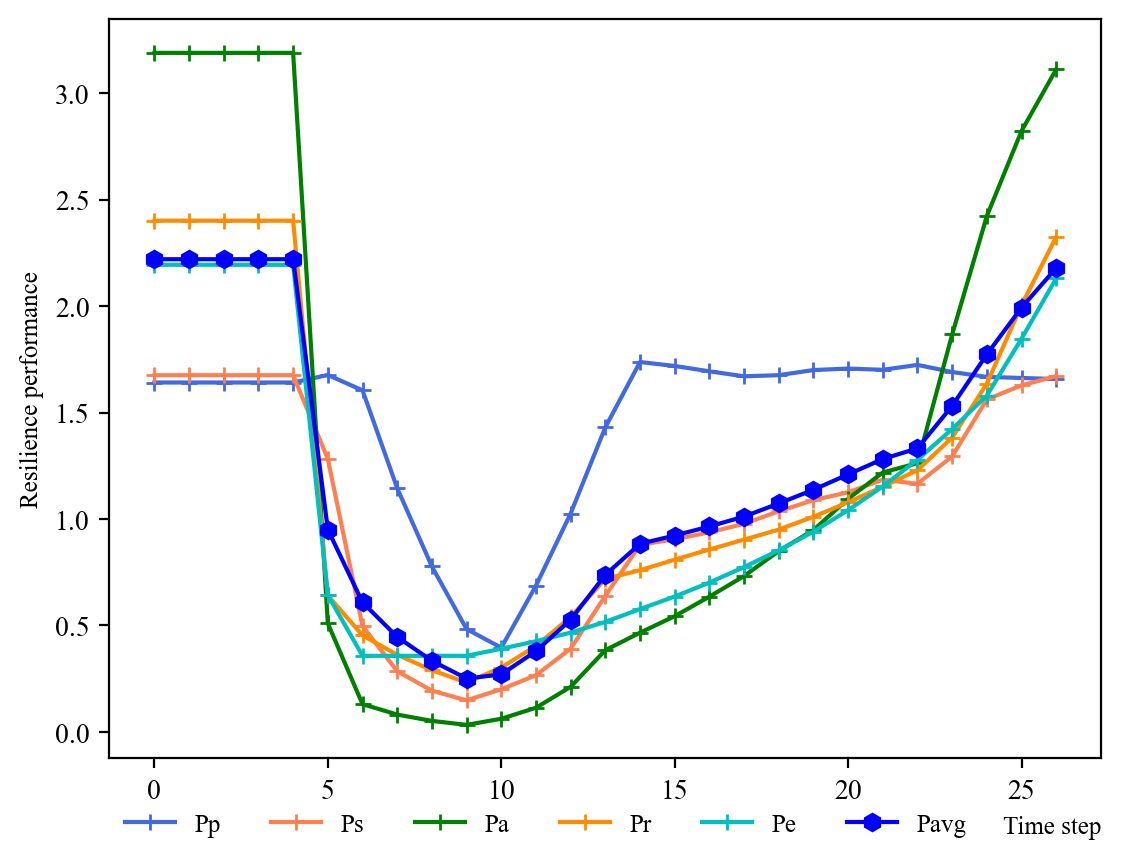}\label{fig:5b}
		\end{minipage}
	}
	\centering
	\caption{The time-varying process of resilience capabilities and performances in ER network under random attack, from Fig.4(e)}
	\label{fig5}
\end{figure*}

There has been significant research work performed to evaluate network resilience in recent years. Xu et al. \cite{ref31} used the relative network size $ G(t) = N(t)/N_{0} $ as the resilience performance $ P(t) $, where $ N(t) $ is the size of the network's largest connected component during recovery process. Moreover, Alenazi et al. \cite{ref15}\cite{ref32} employed networking flow robustness as the measure indicator of network resilience, which is described as formula (4) in section III-A. To simplify the description and comparison in the following experiment, we named these two evaluating methods as \textit{compare1} and \textit{compare2} respectively. The purpose of this experiment is to demonstrate that our proposed evaluation approach perform better evaluating effects than other methods. The experiment will be performed in certain configurations and assumptions. We make the convention and assumption that the network's destruction mainly focuses on node failure, and once one node fails, the connected edges will fail as well. Then, the failed nodes and edges will be removed from the original network. Moreover, two types of simulated attack behaviors, random-based attack and centrality-based attack, are conducted in the experiment. The random-based attack will randomly delete a given number of nodes and their connected edges from the original network graph. Contrastingly, the centrality-based attack will destroy nodes with larger degree and their connected edges from the network. The relative parameters of experiment are listed in Table \ref{tab:table1}.
\begin{table}[!tp]
	\centering
	\caption{\label{tab:table1}Parameters set in comparison experiment}
	\begin{tabular}{lll}
		\toprule
		Symbol & Value & Description \\
		\midrule
		$ T_{d} $ & 5 & The attack occurs at time step 5 \\
		$ N_{d} $ & $ N / 3 $ & The number of deleted nodes during destruction \\
		$ N_{r} $ & $ 2 $ & The number of recovered nodes in each time step \\
		$ PA $ & 1 & The probability of effective attack\\
		$ PR $ & 1 & The probability of recovery of nodes and edges \\
		\bottomrule
	\end{tabular}
\end{table}

\begin{figure*}[hbp]
	\centering
	
	\subfigure[BA network under three levels of attack]{
		\begin{minipage}[t]{0.31\linewidth}
			\centering
			\includegraphics[scale=.4]{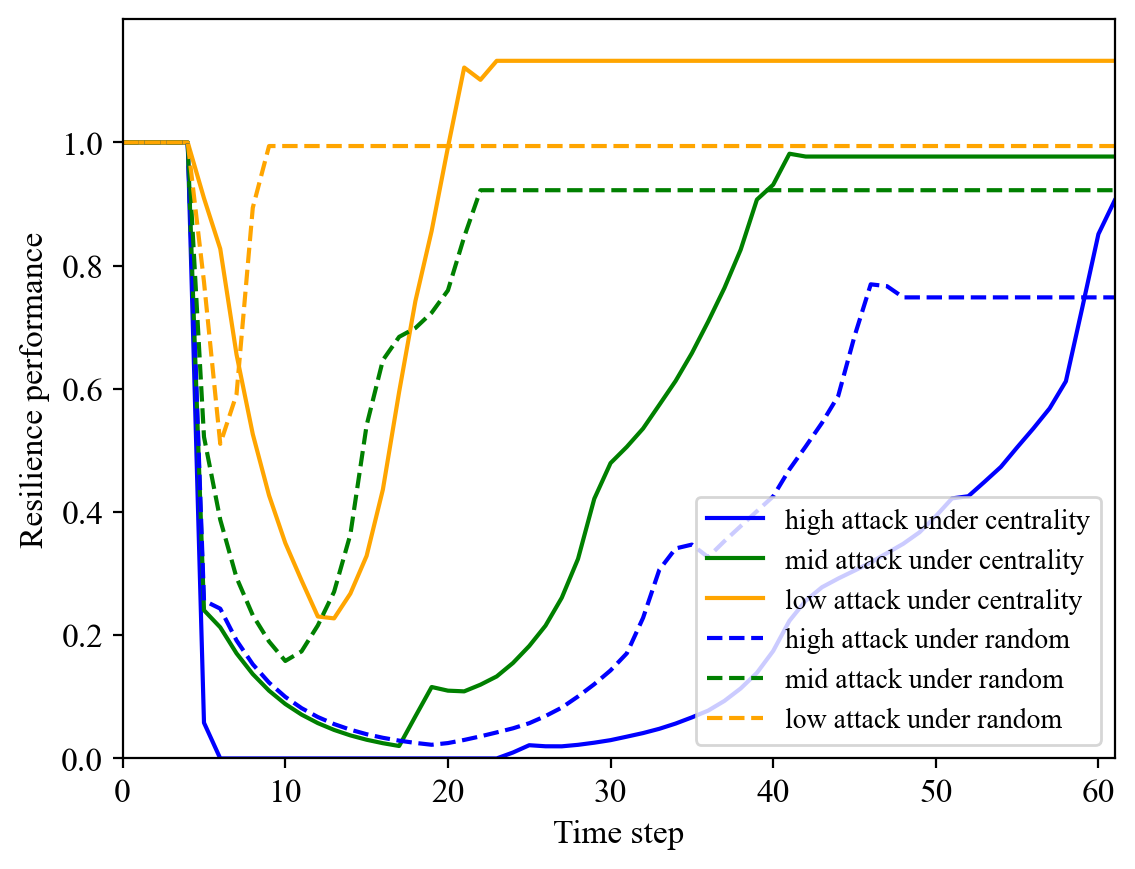}\label{fig:6a}
		\end{minipage}
	}
	\subfigure[ER network under three levels of attack]{
		\begin{minipage}[t]{0.31\linewidth}
			\centering
			\includegraphics[scale=.4]{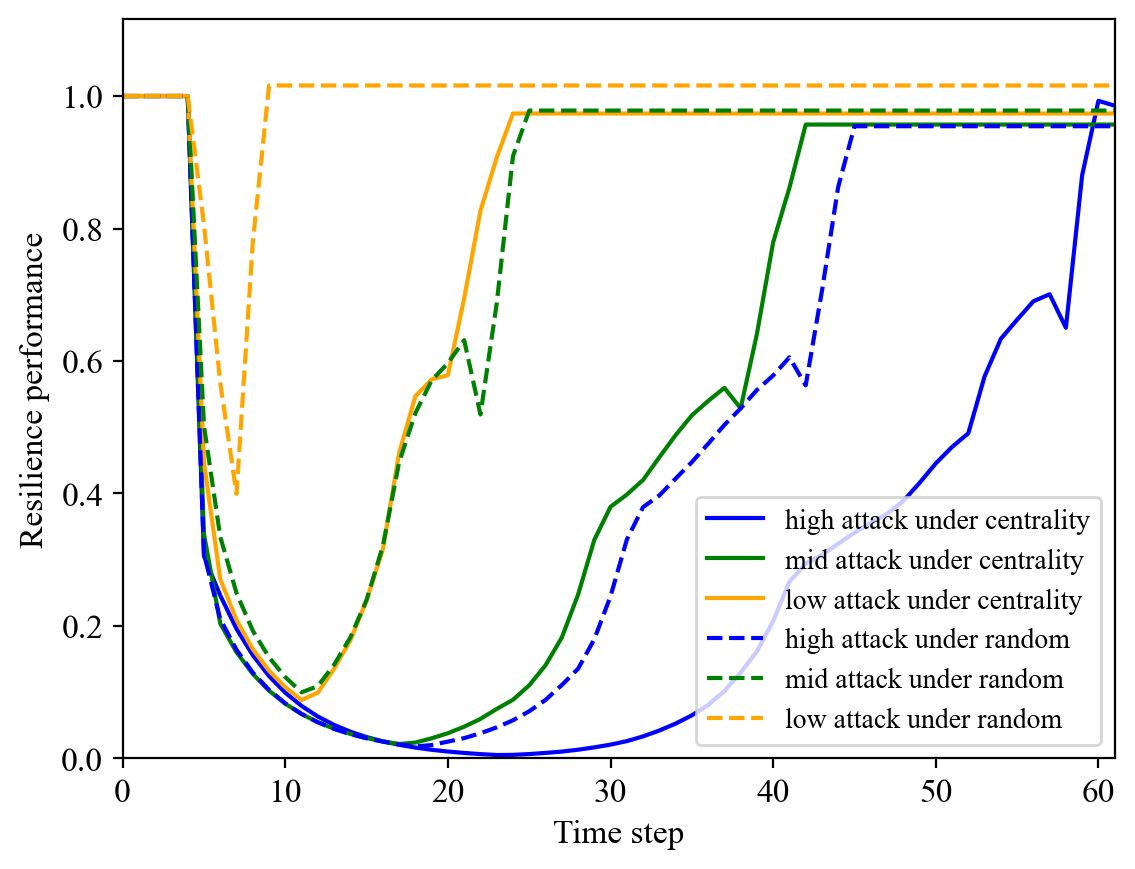}\label{fig:6b}
		\end{minipage}
	}
	\subfigure[Geant2012 under three levels of attack]{
		\begin{minipage}[t]{0.31\linewidth}
			\centering
			\includegraphics[scale=.4]{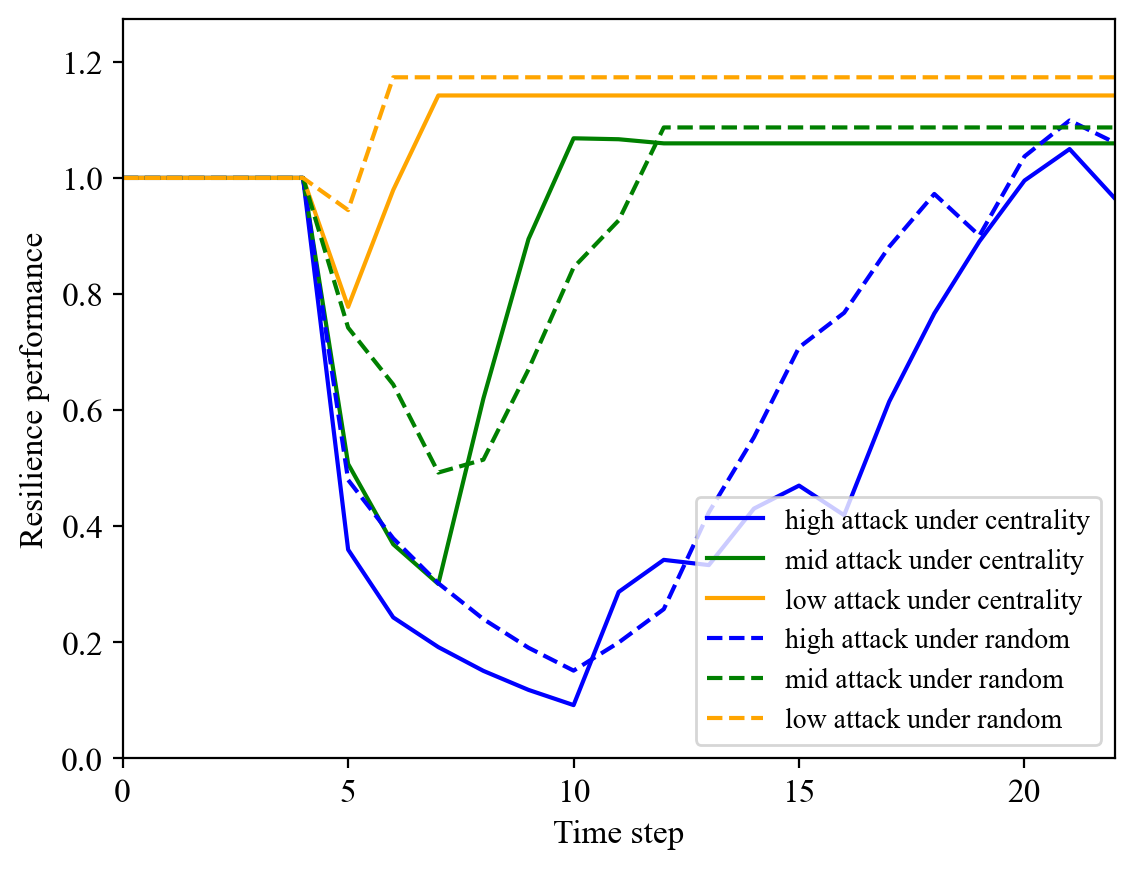}\label{fig:6c}
		\end{minipage}
	}
	
	\subfigure[BA network under three levels of recovery]{
		\begin{minipage}[t]{0.31\linewidth}
			\centering
			\includegraphics[scale=.4]{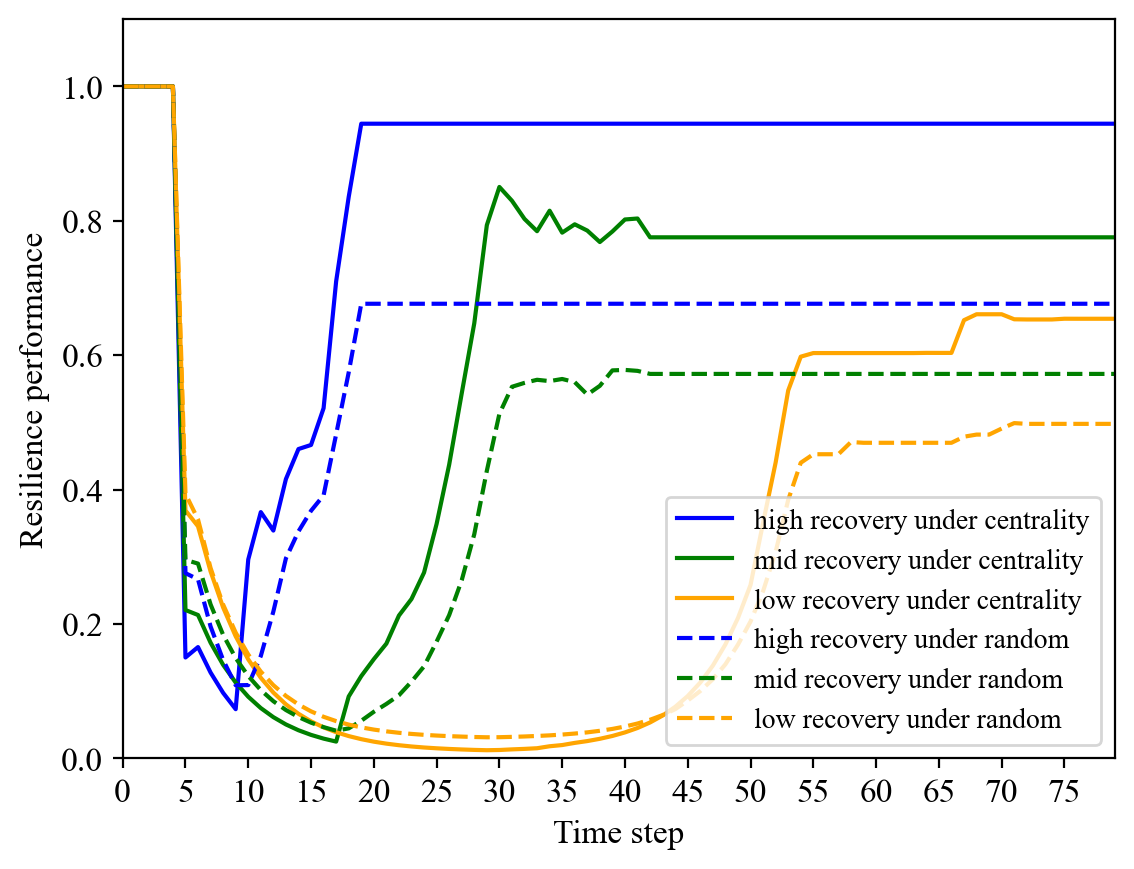}\label{fig:6d}
		\end{minipage}
	}
	\subfigure[ER network under three levels of recovery]{
		\begin{minipage}[t]{0.31\linewidth}
			\centering
			\includegraphics[scale=.4]{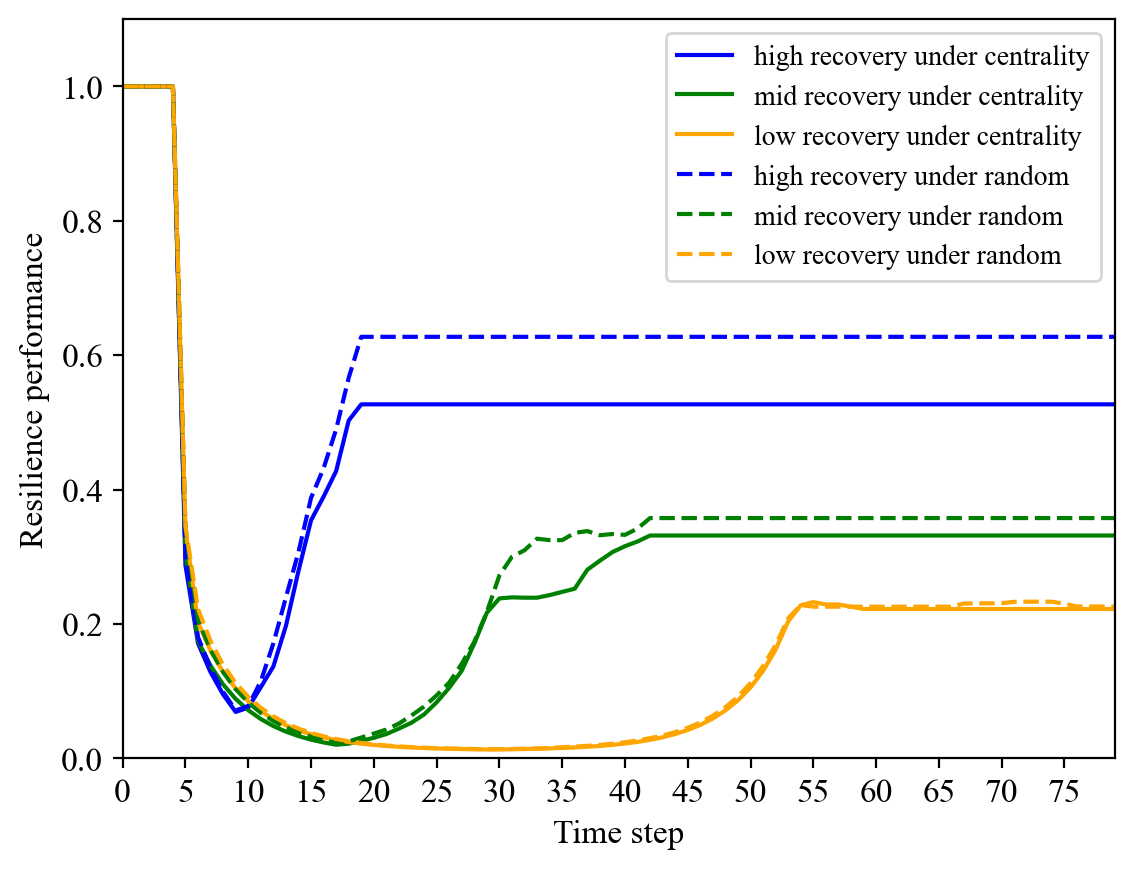}\label{fig:6e}
		\end{minipage}
	}
	\subfigure[Geant2012 under three levels of recovery]{
		\begin{minipage}[t]{0.31\linewidth}
			\centering
			\includegraphics[scale=.4]{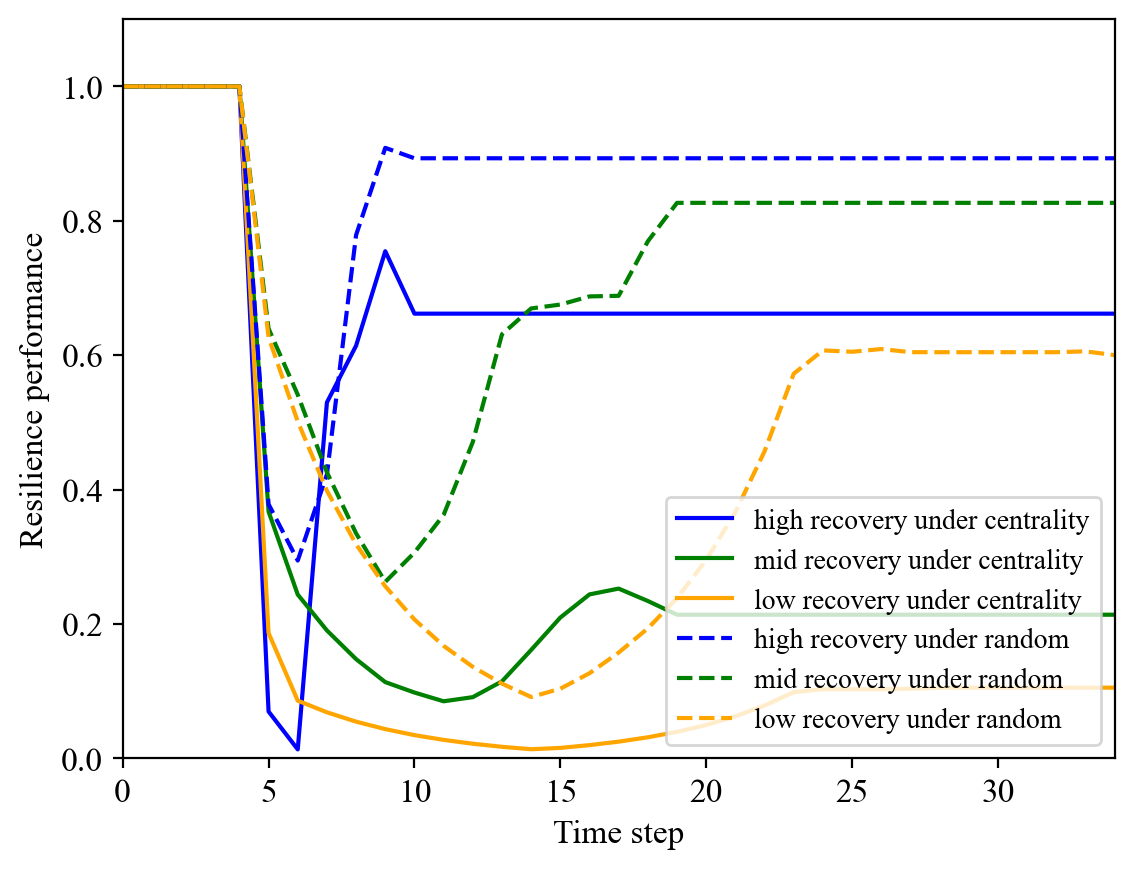}\label{fig:6f}
		\end{minipage}
	}
	
	\centering
	\caption{Comparison among different levels of attack and recovery intensity in three networks and two changing patterns}
	\label{fig6}
\end{figure*}

The results are illustrated in Fig.\ref{fig4}, and the horizontal axis represents the running time step, which records the network's transformation from destruction to recovery, and the vertical axis represents the performance of network resilience as described in formula (17). The blue lines represent our proposed resilience evaluation performance under two types of attacks in three networks. The green lines and black lines represent the \textit{compare1} and \textit{compare2} evaluation methods respectively. After the attack occurred at time step 5 in the BA and ER networks, the resilience performances  declined due to the removal of nodes and edges. The performance of \textit{compare1} and \textit{compare2} maintain the invariable until the network launches recovery strategies. In contrast, the performance of proposed method continues to decline after destruction. The reason for these differences is that the measurement indicator of \textit{compare1} only depends on the network connectivity, or it can be said that it only focuses on the number of nodes in the maximum connected sub-graph, ignoring the network links and capacities. Similarly, taking the network's flow into consideration, the measurement of \textit{compare2} (network flow robustness) shows lower performance reduction due to relatively higher accuracy than \textit{compare1}. Both of these record only the network's structural properties, which is a relatively constant state value. The evaluation results of \textit{compare1} and \textit{compare2} neglect many network attributes and functions, which are far from consistent with the real network situation. In contrast, the proposed method considers both the static network connection states and dynamic capacities of nodes and edges. The time-varying transitive resilience performance evaluation network is constructed based on the DBN, so it can reflect the realistic network resilience in a more detailed and real manner. Besides, as shown in Fig.\ref{fig4}, the resilience performance of \textit{compare1} and \textit{compare2} sharply drops to very low levels when suffering an attack, because the real Internet topology of Geant2012 is less robust than the generated networks. However, the proposed approach does not reduce the evaluated resilience to a lower level by taking into account other network's key factors. Compared with existing approaches, the proposed approach obtains higher resilience performance in Fig.\ref{fig4}, In Fig.\ref{fig:4c}, the final value of \textit{proposed}, \textit{compare1}, \textit{compare2} is 0.95, 0.72, 0.56, respectively. Compared with \textit{compare1} and \textit{compare2}, the evaluation performance of \textit{proposed} are improved $ 31.9\% $ and $ 69.6\% $ respectively. In Fig.\ref{fig:4f}, the final value of \textit{proposed}, \textit{compare1}, \textit{compare2} is 1.13, 1.00, 1.00, respectively. Compared with \textit{compare1} and \textit{compare2}, the evaluation performance of \textit{proposed} are improved $ 13.0\% $. In order to illustrate the intermediate variation process of resilience capabilities and basic performances under our proposed evaluation method, we displayed the detailed capabilities and performances of the proposed method of Fig.\ref{fig:4e} (ER network under random attack). Fig.\ref{fig:5a} shows the variation process of five core capabilities, and Fig.\ref{fig:5b} shows the variation process of five basic resilience performances and their average value, which are calculated by the DBN from five core capabilities. The average value of five basic performances ($ P_{avg} $) in Fig.\ref{fig:5b} represents the blue line ($ proposed $) in  Fig.\ref{fig:4e} before normalization.

\subsection{Evaluation under different attack and recovery scenarios}
The key point of this paper is to effectively evaluate the network resilience in the process of dynamic change, rather than specific defense strategies and recovery algorithms. Therefore, the network's attacks and the recovery strategies become the input variables for our evaluation framework. In order to verify the proposed evaluation model's performance under different attack and recovery scenarios, we conducted two group of controlled experiments in scenarios for three levels of attack intensity and three levels of recovery intensity. Each scenario contains three types of networks and two types of network changed patterns, which include random-based and centrality-based attack and recovery patterns. The controlled parameters are listed in Table \ref{tab:table2}. In the changed attack intensity scenarios, the high level attack is set as $ N_{d} = (3/4)N $ and $ PA = 0.8 $, which represent that the $ 3/4 $ of original network nodes will be affected by the attack and a part of these nodes whose $ L_{i} > (1-PA) $ will be compromised to the attack. Meanwhile, the number of recovered nodes in each time step is set as 2, and the probability of recovery of nodes and edges $ PR $ is set as 1. The purpose of $ PR=1 $ is to make the controlled experiment more unified in recovery stage to demonstrate the differences between changed attack intensity scenarios. Similarly, in the changed recovery intensity scenarios, the number of deleted nodes during destruction is set as $ (1/2)N $ and $ PA = 1 $, which means that a fixed number of nodes will be deleted at attack time. These can also eliminate the influence of different attacks on evaluation of changed network recovery strategies.
\begin{table}[tp]
	\centering
	\caption{\label{tab:table2}Parameters set in changed experiment}
	\begin{tabular}{cc|cccc}
		\toprule
		Scenario & Level & $ N_{d} $            & $ PA $     & $ N_{r} $          & $ PR $    \\ \midrule
		\multirow{3}{*}{\begin{tabular}[c]{@{}c@{}}changed\\ attack\end{tabular}}   & high  & (3/4)$ N $                  & 0.8                & \multirow{3}{*}{2} & \multirow{3}{*}{1} \\
		& mid   & (1/2)$ N $                  & 0.5                &                    &                    \\
		& low   & (1/4)$ N $                  & 0.2                &                    &                    \\ \midrule
		\multirow{3}{*}{\begin{tabular}[c]{@{}c@{}}changed\\ recovery\end{tabular}} & high  & \multirow{3}{*}{(1/2)$ N $} & \multirow{3}{*}{1} & 5                  & 0.8                \\
		& mid   &                      &                    & 2                  & 0.5                \\
		& low   &                      &                    & 1                  & 0.2                \\ \bottomrule
	\end{tabular}
\end{table}
The results of second experiment are displayed in Fig.\ref{fig6}. The Fig.\ref{fig:6a} shows the resilience performance under changed attack intensity scenario in BA network. With the increase of attack intensity, the network's resilience performance will be reduced to a lower level in the resistance stage. As the blue solid curve shown, the resilience performance reaches lowest level after time 5 under high intensity of attack and the centrality-based changed pattern. Compared with the blue dotted line, the centrality-based network destroyed pattern preferentially removes nodes with larger degree, resulting in the faster declined rate of resilience performance. This feature is especially obvious in BA network with centrality characteristic (a small number of nodes connect with numerous edges). Meanwhile, the centrality-based network destroyed pattern will cost more time to rebuild nodes and edges during the recovery stage. The network can achieve a higher level of recovery under low-attack intensity through the evolution stage. The evolution phenomenon in simulation can be implemented by deep reinforcement learning (DRL) in network area. The DRL algorithm can learn about the experience of network strategies and resilience value from previous stages to make more intelligent decision, such as routing optimization \cite{sun2020}, flow migration \cite{xu2020}, and redistribution of resources. These will result in a higher level of network resilience. As the most significant feature of resilience, evolution ability is bound to arouse widespread attention in the future. As is shown in Fig.\ref{fig:6b}, different from Fig.\ref{fig:6a}, except for the minimum level of descent and recovery time, there exists no significant difference in declination rate and last recovery level between changed attack intensity. Because the connection of nodes and edges is randomly generated, and the distribution of nodes' degree is subject to the Poisson distribution, which suggests that different attack intensities and destroyed patterns will not significantly influence resilience performance. When Fig.\ref{fig:6c} is analyzed, the general characteristics above can also be proved in a real-world network Geant2012.

Let's turn our attention to the effect on network resilience performance of different recovery intensities. Due to the same attack intensity parameters at attack time, the network resilience performance exhibits similar decline during the resistance stage. On the whole, the high intensity recovery results in faster recovery speed and higher last recovery performance level than middle and low intensity. However, there are still some differences caused by recovery pattern among the different networks. In Fig.\ref{fig:6d}, BA network shows a slightly faster recovery rate and reaches a higher recovery level in the centrality-based recovery pattern. The reason is that the centrality-based recovery pattern is more suitable for the BA network with a same centrality-based topology. Similarly, under each recovery intensity, the random-based recovery pattern obtains a relatively higher recovery level in Fig.\ref{fig:6e}. When analyzing Fig.\ref{fig:6f}, the general characteristics above can also be proven in real-world network Geant2012. Under the same recovery pattern, curve with a high recovery intensity can give rise to a higher last recovery level.

The above two experiments prove that the proposed network resilience evaluation algorithm has a better evaluating effect than the compared existing works. The complete resilience process including preparation, resistance, adaptation, recovery and evolution stages can be well characterized by our proposed resilience evaluation model. Meanwhile, the resilience evaluation framework can be applied to various attack and recovery scenarios in different networks, which equips considerable applicability and universality.

\section{Related Works}
In general, the resilience evaluation approaches can be mainly divided into two major directions: qualitative and quantitative. The former qualitative approach, which includes methods for assessing system resilience in the absence of numerical descriptors, contains two subcategories: (i) a conceptual framework that provides best practices, and (ii) semi-quantitative indicators that provide expert assessment of different qualitative aspects of resilience. While the latter quantitative approach focuses on general metrics which includes two subcategories: (i) general resilience methods that provide probability or deterministic metrics to quantify across applications, and (ii) structure-based modeling methods that model the domain-specific representations of resilience components \cite{ref2}. It should be noted that the focus of this paper is on quantitative approaches in the computer network domain.

There has been some research work performed on measuring or evaluating network resilience in recent years. Ahmadian \cite{ref12} proposed a method to quantitatively measure network resilience for general physical networked systems, and defined network resilience as a function of criticality, disruption frequency, disruption impact, and recovery capability. Wang \cite{ref13} established the mapping relationship between the physical network and the logical network in the IoT system to build the cascade failure model, and the critical network failure probability and the cascade length are employed as resilience metrics of the IoT system. Zhang \cite{ref14} measured network resilience through formulating and iterating transfer matrix of networking nodes and edges. However, only considering the effects of nodes and edges is not suitable for actual complex networks, and more impact indicators and larger scale network simulation experiments should be considered in the future. Alenazi \cite{ref15} investigated a set of graph spectral robustness metrics and evaluated their accuracy in predicating network resilience against three centrality-based node attacks. Although it was comprehensively considered  from the perspective of graph spectral metrics, the measurement model did not reflect the time-varying feature of the dynamic practical networked system. Xu \cite{ref31} researched the impact of recovery resources allocation approaches on the recovery of scale-free networks under the constraint of a fixed amount of total resources. But the measurement of resilience performance was simply calculated by relative network size, which cannot reflect the true resilience ability of complex networked systems. Yodo \cite{ref16} proposed the DBN approach as a readily available modeling tool to quantify resilient performance in engineered resilient systems, which can capture the dynamic behavior of system performance, but only considering reliability and restoration as the quantification metrics. The existing research's shortcoming is a lack of the systematic view of network resilience, and a lack of combined processes, such as comprehensive resilience capabilities, metric indicators, and time-varying resilience changing processes.

\section{Conclusion and Future Work}
Evaluating network resilience against random failures and target attacks is an important work in network evaluation and defense. In this paper, a comprehensive framework was proposed to qualify the network's time-varying resilience. This framework was developed based on the definition of the dynamic capacities of network components and the measurement of five proposed core network resilience capabilities, which are suitable for the multi-stage processes of network resilience. The DBN approach was employed to quantify the five fundamental and crucial indicators of network resilience performance in temporal network. The simulation experiments were developed to validate the effectiveness and universality of proposed evaluation framework. For future work, we plan to build an resilience evaluation system in physical network environment. Additionally, SDN and networking slice technology deserve attention for providing resilient network capability in future research.

\appendices
\section*{Acknowledgement}
This work is supported by National Key Research and Development Program of China under Grant No. 2018YFB1800602 and No. 2017YFB0801703, CERNET Innovation Project under Grant No. NGIICS20190101 and No. NGII20170406, and National Natural Science Foundation of China (61602114).

\bibliographystyle{unsrt}
\bibliography{refer}

\end{document}